\UseRawInputEncoding

\RequirePackage{fix-cm}
\documentclass{svjour3}                     

\smartqed  

\usepackage{graphicx}

\usepackage{breakcites}
\usepackage[misc]{ifsym}     
\usepackage{siunitx}         
\usepackage[numbers,sort&compress]{natbib}          
\usepackage{amssymb}          
\usepackage{multirow}        
\usepackage[justification=centering]{caption}    
\usepackage{verbatimbox}    
\usepackage{algorithm}  
\usepackage{float}
\usepackage{algpseudocode}  

\usepackage{url}

\begin{document}

\title{Formal modeling and performance evaluation for hybrid systems$\colon$ a probabilistic hybrid process algebra-based approach	
}
\author{Fujun Wang\textsuperscript{1}         \and
        Zining Cao\textsuperscript{1,2,3,4,5}  \and  
        Lixing Tan\textsuperscript{1}     \and
        Zhen Li\textsuperscript{1}
}

\institute{%
	\begin{itemize}
		\item[\textsuperscript{\Letter}] {Zining Cao} \newline
		cstnuaa@gmail.com
		\\
		\\
		{Fujun Wang} \newline
		wangfj@nuaa.edu.cn
		\at
		\item[\textsuperscript{1}] College of Computer Science and Technology, Nanjing University of Aeronautics and Astronautics, Nanjing, China 
		\newline
		\item[\textsuperscript{2}] Key Laboratory of Safety-Critical Software, Nanjing University of Aeronautics and Astronautics, Ministry of Industry and Information Technology, Nanjing, China
		\newline
		\item[\textsuperscript{3}] MIIT Key Laboratory of Pattern Analysis and Machine Intelligence, Nanjing University of Aeronautics and Astronautics, Nanjing, China
		\newline
		\item[\textsuperscript{4}] Science and Technology on Electro-optic Control Laboratory, Luoyang, China
		\newline
		\item[\textsuperscript{5}] Collaborative Innovation Center of Novel Software Technology and Industrialization, Nanjing, China
	\end{itemize}
}            

\date{Received: date / Accepted: date}
\maketitle

\begin{abstract}
Probabilistic behavior is omnipresent in computer controlled systems, in particular, so-called safety-critical hybrid systems, because of various reasons, like uncertain environments, or fundamental properties of nature. In this paper, we extend existing hybrid process algebra ACP$_{hs}^{srt}$ with probability without replacing nondeterministic choice operator. In view of some shortcomings in existing approximate probabilistic bisimulation, we relax the constrains and propose a novel approximate probabilistic bisimulation relation. After that, we present a performance evaluation language CTRML, to reason over probabilistic systems, which extend the results to real number. Along with the specification language, we present a set of algorithms for the evaluation of the language. Additionally, we transfer the hybrid process algebra to probabilistic transition system and show experimental results.

\end{abstract}

\section{Introduction}
\label{intro}
Hybrid process algebras such as hybird Chi \cite{van2006syntax}, HCSP  \cite{chaochen1995formal,zhan2013formal}, HyPA \cite{cuijpers2005hybrid}, $\phi$-calculus \cite{rounds2003the} and ACP$^{srt}_{hs}$ \cite{bergstra2005process}, are well-established techniques for modelling and reasoning about functional aspects of hybrid systems \cite{khadim2006comparative}. The motivation for extending quantified information (like probability) to hybrid process algebras is to develop techniques dealing with non-functional aspects of process behaviors, such as performance and reliability. But in the real-life systems, not only the functionality but also quantitative aspects of the system behaviors are important. We may want to investigate, e.g., the average response time of a system, the probability that a certain failure occurs, or the number of retransmissions that should be done in order to have the message delivered correctly. An analysis of these and similar properties requires that some form of information about the stochastic distribution or the probability over the occurrence of relevant events are put into the model. For instance, performance evaluation is often based on modeling a system as a continuous-time Markov process, in which distributions over delays between actions and over the choice between different actions are specified. Similarly, reliability can be analyzed quantitatively only if we know some probability of the occurrence of events related to a failure \cite{jonsson2001probabilistic}. In this paper, we will adopt probabilistic transition system (PTS) with observation as the basic model for studying. To our knowledge, there exists many hybrid process algebras \cite{bergstra2005process,cuijpers2005hybrid,rounds2003the,van2006syntax,chaochen1995formal,zhan2013formal} or hybrid process calculus \cite{cao2013on}, however, only HCSP has been extended with probability and stochasticity \cite{peng2015extending}. To this end, non-deterministic choice is replaced by probabilistic choice in their method. Till now, there are mainly two approaches to extend process calculus or process algebra with probabilities.  One approach is to replace alternative composition by probabilistic choice. In this case, a fully probabilistic model of a system is obtained. On the other hand, some models allow probabilistic choice as well as alternative composition. In the first approach, non-deterministic choice can be obtained from probabilistic choice, (e.g., for P$+_1$Q we can get process P with probability 1, and for P$+_0$Q we can get process Q, if $+_p$ is the probability choice operator with $p$ in [0, 1]). In many practical circumstances, non-deterministic choice and probabilistic choice are co-exists. In other words, in the presence of a probabilistic choice operator, we still have a need of non-deterministic choice, because (for more detailed reasons we refer the reader to \cite{suzana2002probabilistic}):  

- alternative composition used in the interleaving approach of parallel composition does not model uncertainty but independent activities of the parallel processes or a lack of information for their dependencies;

- alternative composition is very practical in modelling value passing; 

- non-determinism may not make much sense for people doing performance analysis, but in formal methods, the main issue is functionality of systems, (correctness, deadlock-freeness) whether probabilistic aspects are taken into account in the specification of the system or not. 
\\
\indent Due to these reasons, we think that replace non-deterministic choice with probabilistic choice may have some shortcomings. In this paper, we extend hybrid process algebra with probability without replacing non-deterministic, but with some modifications. We do not take the idea of alternative model, which consist of action and probabilistic transitions. Instead, we treat actions to occur with some probability, i.e. to occur simultaneously.
\\
\indent The classical notions of language inclusion, trace equivalence, simulation and bisimulation relations for both discrete and continuous systems are all exact, requiring the behavior of two systems to be identical. When transferring to quantified systems, ideally, for probabilistic simulation, probabilistic bisimulation and so on, quantified information should be identical too. However, when interacting with the physical world, modeled by continuous, discrete or hybrid systems, exact relationships are too restrictive and not robust. For instance, in stochastic systems, the probability values in those models originate either from observations (statistical sampling) or from requirements (probabilistic specification), there is often a bit differences. If we adopt those classical notions like bisimulation (e.g., strong probabilistic bisimulaition \cite{cattani2002decision}, weak probabilistic bisimulaition \cite{lanotte2010weak}) to describe the equivalence of probabilistic processes, then the probability should be matched only when they are identical. In order to bridge the gap between rigid equivalence checking techniques and more relaxed distinguishability oriented requirements of real systems, approximate relationships which explicitly include tiny discrepancies or errors, will be more in line with the actual situations, and much attention has been paid on approximation methods \cite{yan2016approximate,aldini2009note,girard2007approximation}, thus opens up a new research direction. Another motivation is inspired by \cite{aldini2009note}, which takes into account the probability of being in $s$ and in $s^{\prime}$, which are the initial states of the two PTSs under comparison. The initial $\epsilon$-bisimilarity \cite{giacalone1990algebraic} or other approximate probabilistic bisimulation \cite{girard2007approximation} only considers a single-step transition probability difference, ignored the initial probabilistic discrepancy. Instead, their initial distance should somehow receive much more attention.
\\
\indent In this article, we propose a novel approximate probabilistic bisimulation relation, to which the benefits are twofold. One, it takes the initial probability (or distribution) of the two processes into consideration. Two, the constraint on transition probability was modified to a relaxed constraint, which allows for the probabilistic of the transition probability discrepancy beyond a tolerance $\epsilon$ with a confidence at least 1-$\delta$, this modification is more reasonable as the probability of the model may come from observations (statistical sampling), and external environment may have perturbation on the observations so that they may not accurate. At the same time, the value difference of continuous variables is relaxed to within an error range $h$ instead of exact matching. This change is more in line with the actual situation.
\\
\indent Traditional temporal logics, even probabilistic temporal logics are expressive enough, they are limited to producing only true or false responses, as they are still logics. But in real life, performance related queries that cannot be expressed by existing techniques, such as “What is the minimum or maximum probability that the system will reach a failure state within 100 minutes”, this leads us to propose a novel language for performance evaluation. This language can express both performance measures and dependability properties.
\\
\indent The rest of this paper is organized as follows: In Sect. 2, we introduce some preliminary notions used in this paper, like probabilistic bisimulation and probabilistic transition system. In Section 3, we extend hybrid process algebra with probability, and present the related syntax and the transition semantics of it. The approximate probabilistic bisimulation relation and related properties are presented in Sect. 4. The novel language is presented in Sect. 5. In Sect. 6, we present a case study on the nuclear reactor. The transition system of ACP$_{hs}^{srt}$, discretization, algorithm and verification are given in Sect. 7. Finally, in Section 8, we conclude and give some perspectives.

\section{Preliminaries}
\label{sec:1}
In this section, we briefly review some background definitions and notations we will be using in the paper. In what follows, $R_0^{+}$ denote the nonnegative real numbers. In order to measure the discrepancies, we adopt the traditional concept of metric.
\\
\textbf{Definition 1.} (metric \cite{desharnais2002metric}): A metric on a nonempty set $E$ is a function $d: E\times E\rightarrow [0,\infty )$, such that the following three properties hold:\\
1) for all $e_1\in E$, $e_2\in E$, $d(e_1,e_2)=0 \Leftrightarrow e_1=e_2$; \\
2) for all $e_1\in E$, $e_2\in E$, $d(e_1,e_2) = d(e_2,e_1)$; \\
3) for all $e_1\in E$, $e_2\in E$, $e_3\in E$, $d(e_1,e_3)\leqslant d(e_1,e_2)+ d(e_2,e_3)$; \\
then we say that $(E, d)$ is a metric space. If the first property is replaced by $e_1=e_2 \Rightarrow d(e_1,e_2)=0$ then $d$ is call a pseudo-metric. \\
Given a vector $\textbf{x}\in R^n$, \textbf{x} denotes the infinity norm of $\textbf{x}\in R^n$, i.e., $\textbf{x} = max\{|x_1|,|x_2|,$ $|x_3|,...,|x_n|\}$, and define $||\textbf{x}-\textbf{y}||= max\{|x_1-y_1|, |x_2-y_2|, ..., |x_n-y_n|\}$ for any $\textbf{x},\textbf{y}\in R^n$. 
\\
The definition of labeled transition system is given as follows: 
\\
\textbf{Definition2} (Labeled Transition System, LTS \cite{yan2016approximate}): A nondeterministic LTS with observation (without probability) is a tuple $T = < Q, L, \rightarrow, Q_0, Y, H >$, where $Q$ is a set of states, $L$ is a set of labels, $Q_0 \subseteq Q$ is a set of initial states, $Y$ is a set of observations, and $H$ is an observation function $H : Q \rightarrow Y , \rightarrow \subseteq Q \times L \times Q$ is a transition relation.
\\
\indent In this paper, we regard all transition systems with observation as being equipped with metric. In this LTS, we regard $Y$ as being equipped with the metric $d(y_1, y_2) = ||y_1 - y_2||$, unless otherwise specified.
\\
\textbf{Definition 3} (Strong bisimulation): Let $T_i = < Q_i, A, \rightarrow_i, Q_{i}^{0}, Y, H_i> (i=1,2)$ be two LTSs with the same set of actions $A$, observations $Y$ and metric $d$. An equivalence relation $R\subseteq Q_1\times Q_2$ is a strong bisimulation relation between $T_1$ and $T_2$ if for all $(q_1,q_2)\in R$ and for any $a\in A$: \\
1). $d(H_1(q_1),H_2(q_2))=0$ (i.e., $H_1(q_1) = H_2(q_2)$ );\\
2). $\forall q_1\stackrel{a}\longrightarrow_{1}q_1^{\prime},\exists q_2\stackrel{a}\longrightarrow_{2}q_2^{\prime}$ such that $(q_1^{\prime},q_2^{\prime})\in R$ ;\\
3). $\forall q_2\stackrel{a}\longrightarrow_{2}q_2^{\prime}, \exists q_1\stackrel{a}\longrightarrow_{1}q_1^{\prime}$ such that $(q_1^{\prime},q_2^{\prime})\in R$. \\
$T_1$ and $T_2$ are said to be stong bisimilar, if there exists a strong bisimulation relation $R$ between $T_1$ and $T_2$ such that for all $q_1\in Q_1^{0}$, there exists $q_2\in Q_2^{0}$, such that $(q_1,q_2)\in R$, and conversely.\\
As usual, strong bisimilarity, in symbols $\simeq$ , is defined as 
\begin{center}
$\simeq=\cup$\{$R:R$ is a strong bisimulation relation \}.
\end{center}

\indent The strong bisimilarity of two systems is based on the idea of mutual step-by-step simulation. Different from strong bisimulation, weak bisimulation is based on the idea of observation equivalence, which was introduced in the context of nonprobabilistic transition systems in \cite{milner1989communication}. It abstracts away from internal computation by focusing on weak transitions, that is transitions of the form  $\Rightarrow \stackrel{a}\rightarrow \Rightarrow$ (where $\Rightarrow$ is the transitive, reflexive closure of $\stackrel{\tau}\rightarrow$) and requires that weakly bisimilar systems can match each other’s observable behaviors. In other words, whenever a system simulates an action of the other system, it can also execute an arbitrary number (including zero) of internal $τ$ actions before and after the execution of that action. As usual, we define $\hat{a}=\varepsilon$  if $a=\tau$ , otherwise define $\hat{a}=a$.
\\
\textbf{Definition 4 }(Weak bisimulation): Let $T_i = < Q_i, A, \rightarrow_i,Q_{i}^{0}, Y, H_i > (i=1,2)$ be two LTSs with the same set of actions $A$, observations $Y$ and metric $d$. An equivalence relation $R\subseteq Q_1\times Q_2$ is a weak bisimulation relation between $T_1$ and $T_2$ if for all $(q_1,q_2)\in R$ and for any $a\in A$:\\
1). $d(H_1(q_1),H_2(q_2))=0$ (i.e., $H_1(q_1) = H_2(q_2)$ );\\
2). $\forall q_1\stackrel{a}\longrightarrow_{1}q_1^{\prime},\exists q_2\stackrel{\hat{a}}\Rightarrow_{2}q_2^{\prime}$ such that $(q_1^{\prime},q_2^{\prime})\in R$;\\ 
3). $\forall q_2\stackrel{a}\longrightarrow_{2}q_2^{\prime},\exists q_1\stackrel{\hat{a}}\Rightarrow_{1}q_1^{\prime}$ such that $(q_1^{\prime},q_2^{\prime})\in R$.\\ 
$T_1$ and $T_2$ are said to be weak bisimilar, if there exists a weak bisimulation relation $R$ between $T_1$ and $T_2$ such that for all $q_1\in Q_1^{0}$, there exists $q_2\in Q_2^{0}$ such that $(q_1,q_2)\in R$, and conversely.\\
As usual, weak bisimilarity, in symbols $\cong$, is defined as 
\begin{center}
$\cong= \cup$\{$R: R$ is a weak bisimulation relation \}.
\end{center}

\indent In real life, the value of state variables are based on observations (e.g., in cyber-physical systems, the outputs are obtained from sensors), which may be influenced by sensor noise or other perturbations. Constrains of traditional bisimilarity are too restrictive and not robust, and is not applicable to practical applications. Approximate bisimulation provides a robust semantics that is stable with respect to implementation and measurement errors of system behaviors. 
\\
\indent For defining approximate probabilistic weak bisimulation of PTSs, we first give the notion of approximate weak bisimulation of LTSs, which can then be extended to what we needed.
\\
\textbf{Definition 5} ($h$-approximate bisimulation \cite{girard2007approximation}): Let $T_i = < Q_i, A, \rightarrow_i,Q_{i}^{0} , Y, H_i > (i=1,2)$ be two LTSs with the same observations and metric. Let $h\in R_0^{+}$ be the value precision parameter. A symmetric binary relation $R_h \subseteq Q_1\times Q_2$ is an $h$-approximate bisimulation relation between $T_1$ and $T_2$ if for all $(q_1,q_2)\in R_h$ and for any $a\in A$:\\
1). $d(H(q_1),H(q_2))\leqslant h$;\\
2). $\forall q_1\stackrel{a}\longrightarrow_{1}q_1^{\prime},\exists q_2\stackrel{a}\longrightarrow_{2}q_2^{\prime}$ such that $(q_1^{\prime},q_2^{\prime})\in R_h$;\\
3). $\forall q_2\stackrel{a}\longrightarrow_{2}q_2^{\prime},\exists q_1\stackrel{a}\longrightarrow_{1}q_1^{\prime}$ such that $(q_1^{\prime},q_2^{\prime})\in R_h$.\\ 
$T_1$ and $T_2$ are said to be approximately bisimular with precision $h$, if there exists a $h$-approximate bisimulation relation $R_h$ between $T_1$ and $T_2$ such that for all $q_1\in Q_1^{0}$, there exists $q_2\in Q_2^{0}$ such that $(q_1,q_2)\in R_h$, and conversely.\\
As usual, $h$-approximate bisimilarity, in symbols $\simeq_h$, is defined as 
\begin{center}
	$\simeq_h=\cup$ \{$R_h$: $R_h$ is an $h$-approximate bisimulation relation \}.
\end{center}

\textbf{Notation}: Here, we use symbol $h$ instead of $\delta$, as we need to keep consistent with later definitions, and in order to facilitate comparison. 
\\
\indent We adopt the concept of $h$-approximate bisimulation and modify it to $h$-approximate weak bisimulation. In the following, we first give the definition of $h$-approximate weak bisimulation.
\\
\textbf{Definition 6} ($h$-approximate weak bisimulation ): Let $T_i = < Q_i, A, \rightarrow_i,Q_{i}^{0} , Y, H > (i=1,2)$ be two LTSs with the same set of actions, observations and metric. Let $h\in R_0^{+}$ be the value precision parameter. A symmetric binary relation $R_h \subseteq Q_1\times Q_2$ is a $h$-approximate weak bisimulation relation between $T_1$ and $T_2$ if for all $(q_1,q_2)\in R_h$ and for any $a\in A$:\\
1). $d(H_1(q_1),H_2(q_2))\leqslant h$;\\
2). $\forall q_1\stackrel{a}\longrightarrow_{1}q_1^{\prime},\exists q_2\stackrel{\hat{a}}\Rightarrow_{2}q_2^{\prime}$ such that $(q_1^{\prime},q_2^{\prime})\in R$;\\ 
3). $\forall q_2\stackrel{a}\longrightarrow_{2}q_2^{\prime},\exists q_1\stackrel{\hat{a}}\Rightarrow_{1}q_1^{\prime}$ such that $(q_1^{\prime},q_2^{\prime})\in R$.\\ 
$T_1$ and $T_2$ are said to be approximately weak bisimular with precision $h$, if there exists a $h$-approximate weak bisimulation relation $R_h$ between $T_1$ and $T_2$ such that for all $q_1\in Q_1^{0}$,there exists $q_2\in Q_2^{0}$ such that $(q_1,q_2)\in R_h$, and conversely. \\
As usual, weak bisimilarity, in symbols $\cong_h$ , is defined as 
\begin{center}
	$\cong_h= \cup$ \{ $R_h$: $R_h$ is a $h$-approximate weak bisimulation relation \}. 
\end{center}

\indent In the following, the set of actions, denoted by $Act$, is assumed to consist of a set of discrete actions which take no time to execute (written as $\Sigma$), the set of delay actions which just take time delay, and an internal (invisible) action $\tau$. Actions are ranged over $l_1$,$l_2$,…,$l_n$… 
\\
\textbf{Definition 7}(Probabilistic transition system, PTS): A PTS with observation is a tuple $PT= < Q, L, \rightarrow, Q^{0}, Y_1, Y_2, H_1, H_2, H_3 >$, where $Q$ is a finite set of states, $Q^{0} \subseteq Q$ is the set of initial states, $L\subseteq Act$ is a non-empty finite set of actions, $Y_1$ is a set of variables' observations, $Y_2$ is a set of probability observations, and $H_i (i=1,2,3)$ are the observation functions, $H_1$ is an observation function of the values of the state variables $H_1: Q \rightarrow Y_1$, $H_2$ is a function of probabilistic transitions $H_2: Q \times L \times Q \rightarrow [0,1]$, $H_3$ is the probability(distribution) of reaching current states $H_3: Q\rightarrow Y_2$, and $\rightarrow \subseteq Q\times L\times (0,1]\times Q$ is a finite transition relation such that $∀q\in Q$ it holds that $\rm{\Sigma}$$\{ p | a\in Act,t\in S.\ (s, a, p, t)\in\rightarrow \}\in \{0,1\}$ and satisfying the following conditions for continuous transitions:
\\
a).\textbf{identity}: $q\stackrel{0}\rightarrow q$ always holds; \\
b).\textbf{delay determinism}: if $q\stackrel{t}\rightarrow q^{\prime}$ and $q\stackrel{t}\rightarrow q^{\prime\prime}$,then $q^\prime=q^{\prime\prime}$; and \\
c).\textbf{delay additivity}: if $q\stackrel{t_1}\rightarrow q^\prime$ and $q^\prime \stackrel{t_2}\rightarrow q^{\prime\prime}$, then $q\stackrel{t_1+t_2}\longrightarrow q^{\prime\prime}$,where $t,t_1,t_2\in R_0^{+}$. 
\\
In this paper, we modify the definition of PTS, which is not only equipped with the observation information, but also with a transition probability instead of transferring into a distribution on states.
\\
\textbf{Example}: \\
Fig. 1 gives an example of PTS, we only give the detail information about the continuous behavior of the system on state $q_2$, the continuous behavior on other states are similar, we omit it here.
\begin{figure}
	\centering
	\includegraphics[width=0.6\linewidth]{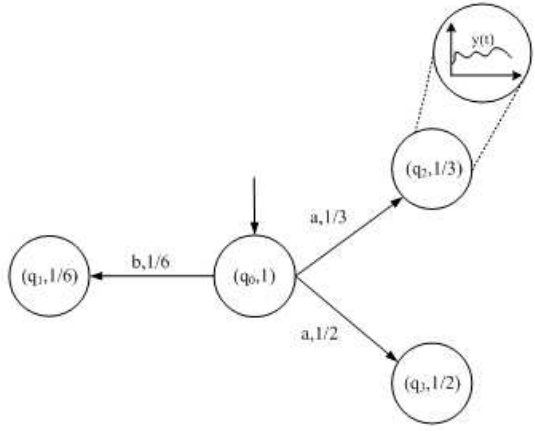}
	\caption{An example of PTS}
	\label{fig:1}
\end{figure}
\\
\textbf{Definition 8 (path)}: A \textit{path} of a PTS \textit{PT} is either a finite or infinite sequence of alternating states and actions $\alpha= s_0a_1s_1a_2s_2…$ starting from a state $s_0$, also denoted by $first(\alpha)$, and, if the sequence is finite, ending with a state, also denoted by $last(\alpha)$, such that for each $i$ > 0 there exists a transition $(s_{i-1},a_i[p_i], s_i)\in \rightarrow$ with $p_i$ > 0.  
\\
\indent We denote by $|c|$ the length of a path $c$, which is the number of occurrences of actions in $c$. If $c$ is infinite, then $|c| = \infty$ . Given a path, $c = s_0a_1s_1a_2s_2…$, the trace of $c$ is denoted by $trace(c)$, is the sub-sequence of external actions of $c$. For instance, for $trace(s_0a_1s_1)= trace(s_0\tau s_1a_1s_2\tau s_3\tau s_4)= a_1$, also denoted by $trace(a_1)$, and $trace(s_0)= trace(s_0\tau s_1\tau s_2\tau s_3\tau s_4)=\epsilon$, the empty sequence, also denoted by $trace(\tau)$.
\\
\indent A configuration of PTS is a pair $(q,\beta,\pi)$, where $q$ is a state of PTS, $\beta$ is a valuation, and $\pi$ is the probability (or distribution) of being in or reaching at state $q$. Given the PTS, we call $S_{PTS}$ the set of configurations of PTS. For the PTSs in this paper, there exist two types of steps, i.e., discrete steps and continuous steps. For a discrete step from a configuration $s_1= (q_1,\beta_1,\pi_1)$ to a configuration $s_2 = (q_2,\beta_2,\pi_2)$ through discrete action $a\in \Sigma\cup\{\tau\}$ with probability $p$, written as $(q_1,\beta_1,\pi_1)\stackrel{a[p]}\longrightarrow (q_2,\beta_2,\pi_2)$ with $\pi_2=\pi_1\times p$. For a continuous step from a configuration $s_1= (q_1,\beta_1,\pi_1)$ to a configuration $s_2 = (q_2,\beta_2,\pi_2)$ through time action $r\in R_0^{+}$ with probability $p$ and state evolution $\rho$, written as $(q_1,\beta_1,\pi_1)\stackrel{r,\rho,[p]}\longmapsto(q_2,\beta_2,\pi_2)$ with $\pi_2=\pi_1\times p$.
\\
\indent For configurations $s_1 = (q_1,\beta_1,\pi_1)$, $s_2 = (q_2,\beta_2,\pi_2)$ and $a\in \Sigma\cup\{\tau\}\cup R_0^{+}$, we define with $P(s_1,a,s_2)$ the probability of reaching configuration $s_2$ from $s_1$ through a transition labelled with $a$. 
\\
\indent A path fragment starting from $s_0$ is a finite sequence of steps $c=s_0\stackrel{a_1}\rightarrow s_1\stackrel{a_2}\rightarrow s_2\stackrel{a_3}\rightarrow \dots \stackrel{a_k}\rightarrow s_k$ such that $s_0,s_1,\dots,s_k\in S_{PTS}, a_1,a_2,…,a_k \in \Sigma\cup\{\tau\}\cup R_0^{+}$ and $\forall i\in \{1,\dots,k\}$  $P(s_{i-1},a_i,s_i)>0$. We define $last(c)=s_k$ and $|c| = k$. If $|c| = 0$ we put $P(c) = 1$, else if $|c|\geqslant 1$, we define $P(c) = P(s_0,a_1,s_1)\cdot \dots\cdot P(s_{k-1},a_k,s_k)$. The set of all paths starting in configuration $s_0$ is denoted by $Path(s_0)$, and the set of finite paths starting in $s_0$ is denoted by $Path_{f}(s_0)$.
\\
\textbf{Definition 9 (Scheduler \cite{probabilistic2012})}: A \textit{scheduler} of paths starting from a configuration $s_0$ and ending in a configuration $s_k$ is a function $\sigma:Path_f(s_0)\longmapsto (\rightarrow \cup \mapsto \cup \{\perp\})$ (where $\perp$ represents \textquotedblleft halt\textquotedblright ) such that for a path $c\in Path(s_0)$, $\sigma(c)$ meets the following two conditions:
\\
1. either $\sigma(c) = last(c)\stackrel{a[p]}\longrightarrow s^\prime$ for some $a\in\Sigma\cup\{\tau\}, s^\prime \in S_{PTS}$ with probability $p$ or $\sigma(c) = last(c)\stackrel{r,\rho,[p]}\longmapsto s^\prime$ for some $a\in R_0^{+}, s^\prime\in S_{PTS}$, state evolution $\rho$ with probability $p$.
\\
2. $\sigma(c) = \perp$, means that the scheduler must stop if the last configuration has reached a deadlock state or a terminating state. 
\\
Once a scheduler is defined and applied on a given PTS, we are not interested in all paths of a PTS, but only in those paths that are obtained after the scheduler is applied on it.

\section{ Extending ACP$_{hs}^{srt}$ with probability (pACP$^{srt}_{hs}$) }
\label{sec:2}
In this section, we introduce our probabilistic extension of process algebra for hybrid systems, which mainly build upon process algebra for hybrid systems \cite{bergstra2005process}. Let us start with some preliminary notations. We use $P, Q$ to denote process, $\psi,\Phi$ for state proposition, $\chi$ for transition proposition, and $H\subseteq A, \alpha \in A$ to denote subsets of actions and an action, respectively.  
\\
\textbf{Syntax}
\\
\indent $P::= \surd \mid \tilde{\tilde{\delta}} \mid \bot \mid \nu_{rel}(P) \mid \sigma_{rel}^r(P) \mid \psi \ ^\blacktriangle \hspace{-0.8em} ^\wedge P \mid \psi:\rightarrow P \mid \phi \  _\blacktriangledown \hspace{-0.9em} ^\bigcap \hspace{0.05em}_V P \mid \chi \ _\blacktriangledown \hspace{-1.2em} \sqcap P \mid P\cdotp P \mid P+P \mid P\parallel P \mid P\mid P \mid P{\parallel \hspace{-0.7em}_\_} P \mid \partial_H(P) \mid \bigoplus p_{i}\tilde{\tilde{ \alpha_{i}}}_{\cdotp}P \mid A$  
\vspace{1ex} \\
The syntax is almost the same as ACP$_{hs}^{srt}$ except $\bigoplus p_{i}\tilde{\tilde{ \alpha_{i}}}_{\cdotp}P $. Here $\bigoplus p_{i}\tilde{\tilde{ \alpha_{i}}}_{\cdotp}P$ stands for a probabilistic choice operator, where $p_{i}$ represents probability, i.e., it satisfies with $p_i \in (0,1]$ and $\Sigma_{i\in 1..n}p_i=1$. When $n=0$, we abbreviate the probabilistic choice as \textbf{0}; when $n=1$, we abbreviate it as $\tilde{\tilde{ \alpha_{1}}}_{\cdotp}P $
\\
\textbf{Structural Operational Semantics rules}
\\
Table 1 Basic rules\\
\rule[5pt]{13cm}{0.05em}\\  
$\frac{<x,\beta,p_0>\stackrel{a[p]}\longrightarrow_{k}<x^{'},\beta^{'},p_1>}{<\sigma_{rel}^0(x),\beta,p_0>\stackrel{a[p]}\longrightarrow_{k}<x^{'},\beta^{'},p_1>}$ \quad
$\frac{<x,\beta,p_0>\stackrel{a[p]}\longrightarrow_{k}<\surd,\beta^{'},p_1>}{<\sigma_{rel}^0(x),\beta,p_0>\stackrel{a[p]}\longrightarrow_{k}<\surd,\beta^{'},p_1>}$ \quad
$\frac{<x,\beta,p_0>\stackrel{r,\rho,[1]}\longmapsto_{k}<x^{'},\beta^{'},p_0>}{<\sigma_{rel}^0(x),\beta,p_0>\stackrel{r,\rho,[1]}\longmapsto_{k}<x^{'},\beta^{'},p_0>}$ \vspace{1ex} \\
$\frac{-}{<\sigma_{rel}^{r+s}(x),\beta,p_0>\stackrel{r,\rho,[1]}\longmapsto_{k}<\sigma_{rel}^s(x),\beta^{'},p_0>}$ \ 
$\frac{\beta^{'}\in[s(x)]}{<\sigma_{rel}^r(x),\beta,p_0>\stackrel{r,\rho,[1]}\longmapsto_{k}<x,\beta^{'},p_0>}$ \ 
$\frac{<x,\beta^{'},p_0>\stackrel{s,\rho\unrhd r,[1]}\longmapsto_{k}<x^{'},\beta^{"},p_0>}{<\sigma_{rel}^r(x),\beta,p_0>\stackrel{r+s,\rho,[1]}\longmapsto_{k}<x^{'},\beta^{"},p_0>}$ \vspace{1ex} \\
$\frac{<x,\beta,p_0>\stackrel{a[p]}\longrightarrow_{k}<x^{'},\beta^{'},p_1>,\beta\in[s(y)]}{<x+y,\beta,p_0>\stackrel{a[p]}\longrightarrow_{k}<x^{'},\beta^{'},p_1>}$ \qquad \qquad \qquad \qquad \qquad
$\frac{\beta\in[s(x)],<y,\beta,p_0>\stackrel{a[p]}\longrightarrow_{k}<y^{'},\beta^{'},p_1>}{<x+y,\beta,p_0>\stackrel{a[p]}\longrightarrow_{k}<y^{'},\beta^{'},p_1>}$ \vspace{1ex} \\ 
$\frac{<x,\beta,p_0>\stackrel{a[p]}\longrightarrow_{k}<\surd,\beta^{'},p_1>,\beta\in[s(y)]}{<x+y,\beta,p_0>\stackrel{a[p]}\longrightarrow_{k}<\surd,\beta^{'},p_1>}$  \qquad \qquad \qquad \qquad \qquad
$\frac{\beta\in[s(x)],<y,\beta,p_0>\stackrel{a[p]}\longrightarrow_{k}<\surd,\beta^{'},p_1>}{<x+y,\beta,p_0>\stackrel{a[p]}\longrightarrow_{k}<\surd,\beta^{'},p_1>}$ \vspace{1ex} \\
$\frac{<x,\beta,p_0>\stackrel{r,\rho,[1]}\longmapsto_{k}<x^{'},\beta^{'},p_0>,<y,\beta,p_0>\stackrel{r}{\not\mapsto}_{k},\beta\in[s(y)]}{<x+y,\beta,p_0>\stackrel{r,\rho,[1]}\longmapsto_{k}<x^{'},\beta^{'},p_0>}$\quad \qquad 
$\frac{<x,\beta,p_0>\stackrel{r}{\not\mapsto}_{k},\beta\in[s(x)],<y,\beta,p_0>\stackrel{r,\rho,[1]}\longmapsto_{k}<y^{'},\beta^{'},p_0>}{<x+y,\beta,p_0>\stackrel{r,\rho,[1]}\longmapsto_{k}<y^{'},\beta^{'},p_0>}$ \vspace{1ex} \\ 
$\frac{<x,\beta,p_0>\stackrel{r,\rho,[1]}\longmapsto_{k}<x^{'},\beta^{'},p_0>,\beta\in[s(x)],<y,\beta,p_0>\stackrel{r,\rho,[1]}\longmapsto_{k}<y^{'},\beta^{'},p_0>}{<x+y,\beta,p_0>\stackrel{r,\rho,[1]}\longmapsto_{k}<x^{'}+y^{'},\beta^{'},p_0>}$ \vspace{1ex} \\
$\frac{<x,\beta,p_0>\stackrel{a[p]}\longrightarrow_{k}<x^{'},\beta^{'},p_1>}{<x\cdot y,\beta,p_0>\stackrel{a[p]}\longrightarrow_{k}<x^{'}\cdot y,\beta^{'},p_1>}$ \qquad \qquad \qquad \qquad \qquad
$\frac{<x,\beta,p_0>\stackrel{a[p]}\longrightarrow_{k}<\surd,\beta^{'},p_1>,\beta^{'}\in[s(y)]}{<x\cdot y,\beta,p_0>\stackrel{a[p]}\longrightarrow_{k}<y,\beta^{'},p_1>}$ \vspace{1ex} \\
$\frac{<x,\beta,p_0>\stackrel{r,\rho,[1]}\longmapsto_{k}<x^{'},\beta^{'},p_0>}{<x\cdot y,\beta,p_0>\stackrel{r,\rho,[1]}\longmapsto_{k}<x^{'}\cdot y,\beta^{'},p_0>}$ \vspace{1ex} \\
$\frac{<x,\beta,p_0>\stackrel{a[p]}\longrightarrow_{k}<x^{'},\beta^{'},p_1>}{<\psi:\rightarrow x,\beta,p_0>\stackrel{a[p]}\longrightarrow_{k}<x^{'},\beta^{'},p_1>}$ $\beta\vDash\psi$ \qquad \qquad \qquad 
$\frac{<x,\beta,p_0>\stackrel{a[p]}\longrightarrow_{k}<\surd,\beta^{'},p_1>}{<\psi:\rightarrow x,\beta,p_0>\stackrel{a[p]}\longrightarrow_{k}<\surd,\beta^{'},p_1>}$ $\beta\vDash\psi$ \vspace{1ex} \\
$\frac{<x,\beta,p_0>\stackrel{r,\rho,[1]}\longmapsto_{k}<x^{'},\beta^{'},p_0>}{<\psi:\rightarrow x,\beta,p_0>\stackrel{r,\rho,[1]}\longmapsto_{k}<x^{'},\beta^{'},p_0>}$ $\beta\vDash\psi$ \vspace{1ex} \\
$\frac{<x,\beta,p_0>\stackrel{a[p]}\longrightarrow_{k}<x^{'},\beta^{'},p_1>}{<\psi \ ^\blacktriangle \hspace{-0.7em} ^\wedge \ x,\beta,p_0>\stackrel{a[p]}\longrightarrow_{k}<x^{'},\beta^{'},p_1>}$ $\beta\vDash\psi$ \qquad \qquad \qquad 
$\frac{<x,\beta,p_0>\stackrel{a[p]}\longrightarrow_{k}<\surd,\beta^{'},p_1>}{<\psi \ ^\blacktriangle \hspace{-0.7em} ^\wedge \ x,\beta,p_0>\stackrel{a[p]}\longrightarrow_{k}<\surd,\beta^{'},p_1>}$ $\beta\vDash\psi$ \vspace{1ex} \\
$\frac{<x,\beta,p_0>\stackrel{r,\rho,[1]}\longmapsto_{k}<x^{'},\beta^{'},p_0>}{<\psi \ ^\blacktriangle \hspace{-0.7em} ^\wedge \ x,\beta,p_0>\stackrel{r,\rho,[1]}\longmapsto_{k}<x^{'},\beta^{'},p_0>}$ $\beta\vDash\psi$ \vspace{1ex} \\
$\frac{<x,\beta,p_0>\stackrel{a[p]}\longrightarrow_{k}<x^{'},\beta^{'},p_1>}{<\phi \  _\blacktriangledown \hspace{-0.7em} ^\bigcap \hspace{0.05em}_V x,\beta,p_0>\stackrel{a[p]}\longrightarrow_{k}<x^{'},\beta^{'},p_1>}$ $\beta\models\phi$ \qquad \qquad \quad
$\frac{<x,\beta,p_0>\stackrel{a[p]}\longrightarrow_{k}<\surd,\beta^{'},p_1>}{<\phi \  _\blacktriangledown \hspace{-0.7em} ^\bigcap \hspace{0.05em}_V x,\beta,p_0>\stackrel{a[p]}\longrightarrow_{k}<\surd,\beta^{'},p_1>}$ $\beta\models\phi$ \vspace{1ex} \\
$\frac{<x,\beta,p_0>\stackrel{r,\rho,[1]}\longmapsto_{k}<x^{'},\beta^{'},p_0>}{<\phi \  _\blacktriangledown \hspace{-0.7em} ^\bigcap \hspace{0.05em}_V x,\beta,p_0>\stackrel{r,\rho,[1]}\longmapsto_{k}<x^{'},\beta^{'},p_0>}$ $\beta\vDash\phi$ \vspace{1ex} \\
$\frac{<x,\beta,p_0>\stackrel{a[p]}\longrightarrow_{k}<x^{'},\beta^{'},p_1>}{<\chi \ _\blacktriangledown \hspace{-0.8em} \sqcap \ x,\beta,p_0>\stackrel{a[p]}\longrightarrow_{k}<x^{'},\beta^{'},p_1>}$ $\beta \rightarrow \beta^{'}\vDash\chi$ \quad \quad
$\frac{<x,\beta,p_0>\stackrel{a[p]}\longrightarrow_{k}<\surd,\beta^{'},p_1>}{<\chi \ _\blacktriangledown \hspace{-0.8em} \sqcap \ x,\beta,p_0>\stackrel{a[p]}\longrightarrow_{k}<\surd,\beta^{'},p_1>}$ $\beta \rightarrow \beta^{'}\vDash\chi$ \vspace{1ex} \\
$\frac{<x,\beta,p_0>\stackrel{r,\rho,[1]}\longmapsto_{k}<x^{'},\beta^{'},p_0>}{<\chi \ _\blacktriangledown \hspace{-0.8em} \sqcap \ x,\beta,p_0>\stackrel{r,\rho,[1]}\longmapsto_{k}<x^{'},\beta^{'},p_0>}$ $\beta \vDash ^{o} \hspace{-0.4em} \chi$ \vspace{1ex} \\
$\frac{<x,\beta,p_0>\stackrel{a[p]}\longrightarrow_{k}<x^{'},\beta^{'},p_1>}{<\nu_{rel}(x),\beta,p_0>\stackrel{a[p]}\longrightarrow_{k}<x^{'},\beta^{'},p_1>}$ \qquad \qquad \qquad \qquad
$\frac{<x,\beta,p_0>\stackrel{a[p]}\longrightarrow_{k}<\surd,\beta^{'},p_1>}{\nu_{rel}(x),\beta,p_0>\stackrel{a[p]}\longrightarrow_{k}<\surd,\beta^{'},p_1>}$ \vspace{1ex}
\\
$\frac{<x,\beta,p_0>\stackrel{a[p]}\longrightarrow_{k}<x^{'},\beta^{'},p_1>}{<A,\beta,p_0>\stackrel{a[p]}\longrightarrow_{k}<x^{'},\beta^{'},p_1>} (A\stackrel{def}= x)$ \qquad \qquad \qquad
$\frac{<x,\beta,p_0>\stackrel{a[p]}\longrightarrow_{k}<\surd,\beta^{'},p_1>}{<A,\beta,p_0>\stackrel{a[p]}\longrightarrow_{k}<\surd,\beta^{'},p_1>}(A\stackrel{def}= x)$ \vspace{1ex}
\\
$\frac{<x,\beta,p_0>\stackrel{r,\rho,[1]}\longmapsto_{k}<x^{'},\beta^{'},p_0>}{A,\beta,p_0>\stackrel{r,\rho,[1]}\longmapsto_{k}<x^{'},\beta^{'},p_0>}(A\stackrel{def}= x)$
\\
\rule[-5pt]{13cm}{0.05em}\\  
\\
\\
Table 2  Rules for $\beta \in [s(\_)]$(r>0) \\
\rule[5pt]{13cm}{0.05em}\\  
$\frac{\beta\in[s(x)]}{\beta\in[s(\sigma_{rel}^0(x))]}$ \qquad 
$\frac{\ }{\beta\in[s(\sigma_{rel}^r(x))]}$ \qquad 
$\frac{\beta\in[s(x)],\beta\in[s(y)]}{\beta\in[s(x+y)]}$ \qquad  
$\frac{\beta\in[s(x)]}{\beta\in[s(x\cdotp y)]}$ \qquad 
$\frac{\beta\in[s(x)]}{\beta\in[s(\psi:\rightarrow x)]}$ \vspace{1ex} \\ 
$\frac{\ }{\beta\in[s(\psi:\rightarrow x)]}\  \beta\nvDash \psi $ \qquad 
$\frac{\beta\in[s(x)]}{\beta\in[s(\psi \ ^\blacktriangle \hspace{-0.7em} ^\wedge \ x)]}\  \beta \vDash \psi $ \qquad 
$\frac{\beta\in[s(x)]}{\beta\in[s(\phi \  _\blacktriangledown \hspace{-0.7em} ^\bigcap \hspace{0.05em}_V x)]}\  \beta \vDash \phi $ \qquad 
$\frac{\beta\in[s(x)]}{\beta\in[s(\chi \ _\blacktriangledown \hspace{-0.8em} \sqcap \ x)]}$ \vspace{1ex} \\ 
$\frac{\ }{\beta\in[s(\chi \ _\blacktriangledown \hspace{-0.8em} \sqcap \ x)]} \beta \nvDash ^{o} \hspace{-0.4em} \chi$ \qquad 
$\frac{\beta\in[s(x)]}{\beta\in[s(\nu_{rel}(x)]}$ \vspace{1ex} \\
$\frac{\beta\in[s(x)],\beta\in[s(y)]}{\beta\in[s(x\parallel y)]}$ \qquad 
$\frac{\beta\in[s(x)],\beta\in[s(y)]}{\beta\in[s(x{\parallel \hspace{-0.3em}_\_} y)]}$ \qquad 
$\frac{\beta\in[s(x)],\beta\in[s(y)]}{\beta\in[s(x\mid y)]}$ \qquad
$\frac{\beta\in[s(x)]}{\beta\in[s(\partial_H(x)]}$ \vspace{1ex} \\
\rule[-5pt]{13cm}{0.05em}\\  
\\
\\
Table 3  Additional Rules for pACP$^{srt}_{hs}$ ($a,b,c\in A,r>0$) \\
\rule[5pt]{13cm}{0.05em}\\  
$\frac{<x,\beta,p_0>\stackrel{a[p]}\longrightarrow_{k}<x^{'},\beta^{'},p_1>,\beta \longrightarrow \beta^{'} \in [d(y)],\beta^{'} \in [s(y)]}{<x\parallel y,\beta,p_0>\stackrel{a[p]}\longrightarrow_{k}<x^{'}\parallel y,\beta^{'},p_1>}$ \quad 
$\frac{\beta \longrightarrow \beta^{'} \in [d(x)],\beta^{'} \in [s(x)],<y,\beta,p_0>\stackrel{a[p]}\longrightarrow_{k}<y^{'},\beta^{'},p_1>}{<x\parallel y,\beta,p_0>\stackrel{a[p]}\longrightarrow_{k}<x\parallel y^{'},\beta^{'},p_1>}$ \vspace{1ex} \\
$\frac{<x,\beta,p_0>\stackrel{a[p]}\longrightarrow_{k}<\surd,\beta^{'},p_1>,\beta \longrightarrow \beta^{'} \in [d(y)],\beta^{'} \in [s(y)]}{<x\parallel y,\beta,p_0>\stackrel{a[p]}\longrightarrow_{k}<y,\beta^{'},p_1>}$ \quad
$\frac{\beta \longrightarrow \beta^{'} \in [d(x)],\beta^{'} \in [s(x)],<y,\beta,p_0>\stackrel{a[p]}\longrightarrow_{k}<\surd,\beta^{'},p_1>}{<x\parallel y,\beta,p_0>\stackrel{a[p]}\longrightarrow_{k}<x,\beta^{'},p_1>}$ \vspace{1ex} \\
$\frac{<x,\beta,p_0>\stackrel{a[p]}\longrightarrow_{k}<x^{'},\beta^{'},p_1>,<y,\beta,p_0>\stackrel{b[q]}\longrightarrow_{j}<y^{'},\beta^{'},p_2>}{<x\parallel y,\beta,p_0>\stackrel{c[p\cdotp q]}\longrightarrow_{(k,j)}<x^{'}\parallel y^{'},\beta^{'},p_3>}\  \gamma(a,b)=c$ \vspace{1ex} \\
$\frac{<x,\beta,p_0>\stackrel{a[p]}\longrightarrow_{k}<x^{'},\beta^{'},p_1>,<y,\beta,p_0>\stackrel{b[q]}\longrightarrow_{j}<\surd,\beta^{'},p_2>}{<x\parallel y,\beta,p_0>\stackrel{c[p\cdotp q]}\longrightarrow_{(k,j)}<x^{'},\beta^{'},p_3>}\  \gamma(a,b)=c$ \vspace{1ex} \\
$\frac{<x,\beta,p_0>\stackrel{a[p]}\longrightarrow_{k}<\surd,\beta^{'},p_1>,<y,\beta,p_0>\stackrel{b[q]}\longrightarrow_{j}<y^{'},\beta^{'},p_2>}{<x\parallel y,\beta,p_0>\stackrel{c[p\cdotp q]}\longrightarrow_{(k,j)}<y^{'},\beta^{'},p_3>}\  \gamma(a,b)=c$ \vspace{1ex} \\
$\frac{<x,\beta,p_0>\stackrel{a[p]}\longrightarrow_{k}<\surd,\beta^{'},p_1>,<y,\beta,p_0>\stackrel{b[q]}\longrightarrow_{j}<\surd,\beta^{'},p_2>}{<x\parallel y,\beta,p_0>\stackrel{c[p\cdotp q]}\longrightarrow_{(k,j)}<\surd,\beta^{'},p_3>}\  \gamma(a,b)=c$ \vspace{1ex} \\
$\frac{<x,\beta,p_0>\stackrel{r,\rho,[1]}\longmapsto_{k}<x^{'},\beta^{'},p_0>,<y,\beta,p_0>\stackrel{r,\rho,[1]}\longmapsto_{k}<y^{'},\beta^{'},p_0>}{<x\parallel y,\beta,p_0>\stackrel{r,\rho,[1]}\longmapsto_{k}<x^{'}\parallel y^{'},\beta^{'},p_0>}$ \vspace{1ex} \\
$\frac{<x,\beta,p_0>\stackrel{a[p]}\longrightarrow_{k}<x^{'},\beta^{'},p_1>,\beta \rightarrow \beta^{'} \in [d(y)],\beta^{'} \in [s(y)]}{<x{\parallel \hspace{-0.3em}_\_} y,\beta,p_0>\stackrel{a[p]}\longrightarrow<x^{'}\parallel y,\beta^{'},p_1>}$ \quad
$\frac{<x,\beta,p_0>\stackrel{a[p]}\longrightarrow_{k}<\surd,\beta^{'},p_1>,\beta \rightarrow \beta^{'} \in [d(y)],\beta^{'} \in [s(y)]}{<x{\parallel \hspace{-0.3em}_\_} y,\beta,p_0>\stackrel{a[p]}\longrightarrow_{k}<y,\beta^{'},p_1>}$\vspace{1ex} \\
$\frac{<x,\beta,p_0>\stackrel{r,\rho,[1]}\longmapsto_{k}<x^{'},\beta^{'},p_0>,<y,\beta,p_0>\stackrel{r,\rho,[1]}\longmapsto_{k}<y^{'},\beta^{'},p_0>}{<x{\parallel \hspace{-0.3em}_\_} y,\beta,p_0>\stackrel{r,\rho,[1]}\longmapsto_{k}<x^{'}{\parallel \hspace{-0.3em}_\_} y^{'},\beta^{'},p_0>}$ \vspace{1ex} \\
$\frac{<x,\beta,p_0>\stackrel{a[p]}\longrightarrow_{k}<x^{'},\beta^{'},p_1>,<y,\beta,p_0>\stackrel{b[q]}\longrightarrow_{j}<y^{'},\beta^{'},p_2>}{<x\mid y,\beta,p_0>\stackrel{c[p\cdotp q]}\longrightarrow_{(k,j)}<x^{'}\parallel y^{'},\beta^{'},p_3>} \ \gamma(a,b)=c$ \vspace{1ex} \\
$\frac{<x,\beta,p_0>\stackrel{a[p]}\longrightarrow_{k}<x^{'},\beta^{'},p_1>,<y,\beta,p_0>\stackrel{b[q]}\longrightarrow_{j}<\surd,\beta^{'},p_2>}{<x\mid y,\beta,p_0>\stackrel{c[p\cdotp q]}\longrightarrow_{(k,j)}<x^{'},\beta^{'},p_3>} \ \gamma(a,b)=c$ \vspace{1ex} \\
$\frac{<x,\beta,p_0>\stackrel{a[p]}\longrightarrow_{k}<\surd,\beta^{'},p_1>,<y,\beta,p_0>\stackrel{b[q]}\longrightarrow_{j}<y^{'},\beta^{'},p_2>}{<x\mid y,\beta,p_0>\stackrel{c[p\cdotp q]}\longrightarrow_{(k,j)}<y^{'},\beta^{'},p_3>} \ \gamma(a,b)=c$ \vspace{1ex} \\
$\frac{<x,\beta,p_0>\stackrel{a[p]}\longrightarrow_{k}<\surd,\beta^{'},p_1>,<y,\beta,p_0>\stackrel{b[q]}\longrightarrow_{j}<\surd,\beta^{'},p_2>}{<x\mid y,\beta,p_0>\stackrel{c[p\cdotp q]}\longrightarrow_{(k,j)}<\surd,\beta^{'},p_3>} \ \gamma(a,b)=c$ \vspace{1ex} \\
$\frac{<x,\beta,p_0>\stackrel{r,\rho,[1]}\longmapsto_{k}<x^{'},\beta^{'},p_0>,<y,\beta,p_0>\stackrel{r,\rho,[1]}\longmapsto_{k}<y^{'},\beta^{'},p_0>}{<x\mid y,\beta,p_0>\stackrel{r,\rho,[1]}\longmapsto_{k}<x^{'}\mid y^{'},\beta^{'},p_0>}$ \vspace{1ex} \\
$\frac{<x,\beta,p_0>\stackrel{a[p]}\longrightarrow_{k}<x^{'},\beta^{'},p_1>}{< \partial_H(x),\beta,p_0>\stackrel{a[p]}\longrightarrow_{k}<\partial_H(x^{'}),\beta^{'},p_1>} \ a \notin H$ \qquad \qquad 
$\frac{<x,\beta,p_0>\stackrel{a[p]}\longrightarrow_{k}<\surd,\beta^{'},p_1>}{< \partial_H(x),\beta,p_0>\stackrel{a[p]}\longrightarrow_{k}<\surd,\beta^{'},p_1>} \ a \notin H$ \vspace{1ex} \\
$\frac{<x,\beta,p_0>\stackrel{r,\rho,[1]}\longmapsto_{k}<x^{'},\beta^{'},p_0>}{< \partial_H(x),\beta,p_0>\stackrel{r,\rho,[1]}\longmapsto_{k}<\partial_H(x^{'}),\beta^{'},p_0>}$ \vspace{1ex} \\
$\frac{<x,\beta,p_0>\stackrel{a[p]}\longrightarrow_{k}<x^{'},\beta^{'},p_1>,\beta \in [s(y)]}{< x\oplus y,\beta,p_0>\stackrel{a[p]}\longrightarrow_{k}<x^{'},\beta^{'},p_1>}$ \qquad \qquad \qquad \qquad
$\frac{<x,\beta,p_0>\stackrel{a[p]}\longrightarrow_{k}<\surd,\beta^{'},p_1>,\beta \in [s(y)]}{< x\oplus y,\beta,p_0>\stackrel{a[p]}\longrightarrow_{k}<\surd,\beta^{'},p_1>}$ \vspace{1ex} \\
\rule[-5pt]{13cm}{0.05em}\\  
\\
\\
Table 4  Rules for $\beta \rightarrow \beta^{'} \in [d(\_)]$ ($ a\in A_\delta,r>0 $) \\
\rule[5pt]{13cm}{0.05em}\\  
$\frac{\beta \rightarrow \beta^{'} \in [d(x)]}{\beta \rightarrow \beta^{'} \in [d(\sigma_{rel}^0(x))]}$ \qquad 
$\frac{\ }{\beta \rightarrow \beta^{'} \in [d(\sigma_{rel}^r(x))]}$ \qquad 
$\frac{\beta \rightarrow \beta^{'} \in[d(x)],\beta \rightarrow \beta^{'} \in[d(y)]}{\beta \rightarrow \beta^{'} \in[d(x+y)]}$ \qquad  
$\frac{\beta \rightarrow \beta^{'} \in [d(x)]}{\beta \rightarrow \beta^{'} \in [d(x\cdotp y)]}$  \vspace{1ex} \\ 
$\frac{\beta \rightarrow \beta^{'} \in [d(x)]}{\beta \rightarrow \beta^{'} \in [d(\psi:\rightarrow x)]}$ \qquad
$\frac{\ }{\beta \rightarrow \beta^{'} \in [d(\psi:\rightarrow x)]}\  \beta\nvDash \psi $ \qquad 
$\frac{\beta \rightarrow \beta^{'} \in [d(x)]}{\beta \rightarrow \beta^{'} \in [d(\psi \ ^\blacktriangle \hspace{-0.7em} ^\wedge \ x)]}\  \beta \vDash \psi $ \vspace{1ex} \\ 
$\frac{\beta \rightarrow \beta^{'} \in [d(x)],<x,\beta,p_0>\stackrel{r,\rho,[1]}\longmapsto<x^{'},\beta^{''},p_0>}{\beta \rightarrow \beta^{'} \in[d(\phi \  _\blacktriangledown \hspace{-0.7em} ^\bigcap \hspace{0.05em}_V x)]}\  \beta \rightarrow \beta^{'} \vDash C_V,\beta\vDash \phi $ \vspace{1ex} \\
$\frac{\beta \rightarrow \beta^{'} \in [d(x)],<x,\beta,p_0>{\not\mapsto}}{\beta \rightarrow \beta^{'} \in [d(\phi \  _\blacktriangledown \hspace{-0.7em} ^\bigcap \hspace{0.05em}_V x)]}\  \beta \vDash \phi $ \quad
$\frac{\beta \rightarrow \beta^{'} \in [d(x)]}{\beta\in[d(\chi \ _\blacktriangledown \hspace{-0.8em} \sqcap \ x)]}$ \quad
$\frac{\ }{\beta \rightarrow \beta^{'} \in [d(\chi \ _\blacktriangledown \hspace{-0.8em} \sqcap \ x)]}\ \beta \nvDash ^{o} \hspace{-0.4em} \chi$ \quad 
$\frac{\beta\in[s(x)]}{\beta \rightarrow \beta^{'} \in [d(\nu_{rel}(x)]}$ \vspace{1ex} \\
$\frac{\beta \rightarrow \beta^{'} \in[d(x)],\beta \rightarrow \beta^{'} \in[d(y)],<x\parallel y,\beta,p_0>\stackrel{r,\rho,[1]}\longmapsto<x^{'},\beta^{"},p_0>}{\beta \rightarrow \beta^{'} \in[d(x\parallel y)]}$ \ \qquad 
$\frac{\beta \in [s(x)],\beta \in [s(y)],<x\parallel y,\beta,p_0>\not\mapsto}{\beta \rightarrow \beta^{'} \in[d(x\parallel y)]}$ \vspace{1ex} \\ 
$\frac{\beta \rightarrow \beta^{'} \in[d(x)],\beta \rightarrow \beta^{'} \in[d(y)],<x{\parallel \hspace{-0.3em}_\_} y,\beta,p_0>\stackrel{r,\rho,[1]}\longmapsto<x^{'},\beta^{"},p_0>}{\beta \rightarrow \beta^{'} \in[d(x{\parallel \hspace{-0.3em}_\_} y)]}$ \qquad
$\frac{\beta \in [s(x)],\beta \in [s(y)],<x{\parallel \hspace{-0.3em}_\_} y,\beta,p_0>\not\mapsto}{\beta \rightarrow \beta^{'} \in[d(x{\parallel \hspace{-0.3em}_\_} y)]}$ \vspace{1ex} \\ 
$\frac{\beta \rightarrow \beta^{'} \in[d(x)],\beta \rightarrow \beta^{'} \in[d(y)],<x\mid y,\beta,p_0>\stackrel{r,\rho,[1]}\longmapsto<x^{'},\beta^{"},p_0>}{\beta \rightarrow \beta^{'} \in[d(x\mid y)]}$ \ \  \qquad
$\frac{\beta \in [s(x)],\beta \in [s(y)],<x\mid y,\beta,p_0>\not\mapsto}{\beta \rightarrow \beta^{'} \in[d(x\mid y)]}$ \vspace{1ex} \\ 
$\frac{\beta \rightarrow \beta^{'} \in[d(x)]}{\beta \rightarrow \beta^{'} \in[d(\partial_H(x)]}$  \qquad
$\frac{\beta \rightarrow \beta^{'} \in[d(x)]}{\beta \rightarrow \beta^{'} \in[d(A)]}(A\stackrel{def}= x) $ 
\\
\rule[-5pt]{13cm}{0.05em}\\  
\\

\section{ Approximate probabilistic bisimulation }
In this section, approximate probabilistic bisimulation relations are the focus of this paper, which fill the gap between theory (i.e., bisimulation relations require the behavior of two systems to be identical) and practice (i.e., due to external factors influence, tiny discrepancies or errors exists between two systems). With the inspiration of \cite{desharnais2002metric}, the probability computation of transitions used in this article is given as follows:
\\
\textbf{Definition 10 }: Let \textit{PT} be a PTS with observation, $q\in Q, E\subseteq Q$. Then, the probability of going from $q$ to $E$ via action $a$, denoted by Pr($q, a, E$), is defined as: 
\begin{center}
	Pr$(q, a, E)=sup\{\sum \limits_{q^\prime\in E}p|q \stackrel{a[p]}\Longrightarrow q^\prime \}$
\end{center}

The supremum in this definition is the source of the subtlety of weak bisimulation--Pr($q, a, \cdotp $) does not satisfy additivity. Additionally, if $E$ is a singleton state, say $q^\prime$, then Pr($q, a, q^\prime$) is the probability of weak transition labeled with $a$ from $q$ to $q^\prime$, defined as: 
\begin{center}
	Pr($q, a, q^\prime$)=$sum\{p|q \stackrel{a[p]}\Longrightarrow q^\prime \}$
\end{center}

\indent As usual, we define $\hat{\alpha}=\epsilon$ if $\alpha=\tau$, otherwise $\hat{\alpha}=\alpha$. 
\\
\indent Similar to \cite{yan2016approximate}, for a maximum sequence of $\tau$ actions $<q_i,\beta_i,\pi_{i}>\stackrel{\tau[p_i]}\longrightarrow <q_{i+1},\beta_{i+1},$ $\pi_{(i+1)}>\stackrel{\tau[p_{i+1}]}\longrightarrow\ldots \stackrel{\tau[p_{i+k-1}]}\longrightarrow <q_{i+k},\beta_{i+k},\pi_{(i+k)}>$, we remove the intermediate states and define the $\tau$-compressed transition $<q_i,\beta_i,\pi_{i}>\stackrel{\tau[p]}\twoheadrightarrow<q_{i+k},\beta_{i+k},\pi_{(i+k)}>$ with $p=p_i\times p_{i+1}\times \ldots \times p_{i+k-1}$ instead. For unification, for a non-$\tau$ transition $<q_i,\beta_i,\pi_{i}>\stackrel{\alpha[p_i]}\longrightarrow <q_{i+1},\beta_{i+1},\pi_{(i+1)}>$ where $\alpha\neq\tau$, we define $<q_i,\beta_i,\pi_{i}>\stackrel{\alpha[p_i]}\twoheadrightarrow <q_{i+1},\beta_{i+1},\pi_{(i+1)}>$. As a common convention in process algebra, we use $<q_i,\beta_i,\pi_{i}>\\ 
\stackrel{\alpha[p_i]}\Longrightarrow <q^\prime,\beta^\prime,\pi_{i}^\prime>$ to denote the closure of $\tau$ transitions, i.e.,$<q_i,\beta_i,\pi_{i}>(\stackrel{\tau[p_l]}\twoheadrightarrow)^{\{0,1\}}$\\$(\stackrel{\alpha[p_m]}\twoheadrightarrow)(\stackrel{\tau[p_n]}\twoheadrightarrow)^{\{0,1\}}<q^\prime,\beta^\prime,\pi_{i}^\prime>$ with $p_i=p_l^{\{0,1\}}\times p_m\times p_n^{\{0,1\}}$, for any $\alpha\in\sigma$ in the sequel. In what follows, we will denote $pTS_i=<Q_i,A_i,\twoheadrightarrow _i,Q^0_i,Y^i_1,Y^i_2,H^i_1,H^i_2,H^i_3>$ the resulting PTS from $<Q_i,A_i,\rightarrow _i,Q^0_i,Y^i_1,Y^i_2,H^i_1,H^i_2,H^i_3>$ by replacing each label transition with itsτ-compressed version.
\\
\textbf{Definition 11:} $(h,\epsilon,\delta)$-approximate probabilistic bisimulation 
\\
Let $pTS_i=<Q_i,A_i,\twoheadrightarrow _i,Q^0_i,Y^i_1,Y^i_2,H^i_1,H^i_2,H^i_3> (i=1,2)$ be two PTSs with the same output sets $Y_1$,$Y_2$, the same metric $d$ and the same set of actions $A$. Let $h\in R_0^{+}$ be the value precision, and $\epsilon,\delta \in [0,1)$ be the probabilistic precisions, respectively. A sysmetric binary relation $\mathcal{B}$ is called an ($h$,$\epsilon$,$\delta$)-approximate probabilistic bisimulation relation between $pTS_1$ and $pTS_2$, if $(<q_1,\beta,p_{01}>,<q_1,\beta,p_{02}>) \in \mathcal{B}$, then \\
1). $d(H^{1}_1(<q_1,\beta,p_{01}>),H^{2}_1(<q_2,\beta,p_{02}>))\leqslant h$;\\
2). $d(H^1_3(<q_1,\beta,p_{01}>),H^2_3(<q_2,\beta,p_{02}>))\leqslant \epsilon$;\\
3). if$<q_1,\beta,p_{01}>\stackrel{\alpha[p_1]}\twoheadrightarrow_1 <q_1^{'},\beta^{'},p_{01}^{'}>$, then there exists $q_2^{'}\in Q_2$, such that $<q_2,\beta,p_{02}>\stackrel{\alpha[p_2]}\Longrightarrow_2<q_2^{'},\beta^{'},p_{02}^{'}>$ with $Pr(\vert p_1-p_2\vert \leqslant \epsilon)\geqslant 1-\delta$ and $(<q_1^{'},\beta^{'},p_{01}^{'}>,\\
<q_2^{'},\beta^{'},p_{02}^{'}>) \in \mathcal{B}$ for any $\alpha \in \Sigma\cup\{\tau\}$;\\
4). if$<q_1,\beta,p_{01}>\stackrel{r,\rho,[1]}\twoheadrightarrow_1 <q_1^{'},\beta^{'},p_{01}^{'}>$, then there exists $q_2^{'} \in Q_2$, such that $<q_2,\beta,p_{02}>\stackrel{r,\rho,[1]}\twoheadrightarrow_2<q_2^{'},\beta^{'},p_{02}^{'}>$ and $(<q_1^{'},\beta^{'},p_{01}^{'}>,<q_2^{'},\beta^{'},p_{02}^{'}>) \in \mathcal{B}$ for any $r \in R^{+}_0$;\\
5). if $\beta \rightarrow \beta^{'} \in[d(q_1)]$, then $\beta \rightarrow \beta^{'} \in[d(q_2)]$.\\
As usual, $(h,\epsilon,\delta)$-approximate probabilistic bisimilarity, in symbols $\cong _{h,\epsilon,\delta}$, is defined as 
\begin{center}
	$\cong _{h,\epsilon,\delta}$ = $\cup$\{$\mathcal{B}:\mathcal{B}$ is an $(h,\epsilon,\delta)$-approximate probabilistic bisimulation relation\}.
\end{center}

\textbf{Notation}: Here, in the above definition, we take the same delay $r$ instead of different delay with tiny discrepancies, because we can treat the delay as execute the same action, so as to simplify this definition.
\\
\textbf{Proposition 1}: Let  $T = < Q, A, \rightarrow, Q^0, Y, H >$ be a nondeterministic LTS. Then \\
(1). $\cong$  is the largest weak bisimulation. \\
(2). $\cong_h$  is the largest $h$-approximate weak bisimulation. \\
(3). $\cong\  \subseteq \ \cong_h$. \\
(4). If $h\leqslant h^\prime$, then $\cong_h\  \subseteq\  \cong_{h^\prime}$ .\\
\textbf{Proof}: \\
(1) and (2) immediately follows from definition 4 and 6, respectively. By definition 6, if $h\leqslant h^\prime$, then we can immediately conclude that every $h$-approximate weak bisimulation is an $h^\prime$-approximate weak bisimulation and, hence, (4) holds. In the following, we show (3). \\
Let $h\in R_0^{+}$, $(q_1,q_2)\in R$ and $R$ is a weak bisimulation relation, then $d(H(q_1),H(q_2))=0$, so $d(H(q_1),H(q_2))\leqslant h$. Assume $q_1\stackrel{a}\rightarrow q_1^\prime$, then there exists $q_2\stackrel{a}\Rightarrow q_2^\prime$ and $(q_1^\prime,q_2^\prime)\in R$ with $\hat{a}=a$ if $a\in \Sigma\backslash \{\tau\}$, and $\hat{a}=\epsilon$ if $a=\tau$,$d(H(q_1^\prime),H(q_2^\prime))=0\leqslant h$, so $(q_1,q_2)\in R_h$. Hence, by definition 6, (3) holds.   
\hfill  $\Box$
\\
\textbf{Proposition 2}: Let $T_1,T_2,T_3$ be three nondeterministic LTSs. Then \\
(1). For all $h\in R_0^{+}$, T$_1\cong_h$ T$_1$. \\
(2). If T$_1\cong_h$ T$_2$ and T$_2\cong_{h^\prime}$ T$_3$, then T$_1\cong_{h+h^\prime}$ T$_3$. \\
\textbf{Proof}: \\
The first property is obvious. So we only need to prove (2).\\
Let $(q_1,q_2)\in R_h, (q_2,q_3)\in R_{h^\prime}$, and $R_h,R_{h^\prime}$ be two approximate weak bisimulation relations, with precision $h$ and $h^\prime$, respectively. Let us define the following relation
$R_{h+h^\prime}$=\{$(q_1,q_3)|\exists q_2$ such that $(q_1,q_2)\in R_h$ and $(q_1,q_3)\in R_{h^\prime}$ \}.
So we only need to prove that $R_{h+h^\prime}$ is an approximate weak bisimulation.
Let $(q_1,q_3)\in R_{h+h^\prime}$, let $q_2$ be the corresponding element of $Q_2$.
$d(H_1(q_1),H_3(q_3))\leqslant d(H_1(q_1),H_2(q_2))+ d(H_2(q_2),H_3(q_3))\leqslant h+h^\prime$.
For all $q_1\stackrel{a}\Rightarrow_1 q_1^\prime$ , there exists $q_2\stackrel{a}\Rightarrow_2 q_2^\prime$  such that $(q_1^\prime,q_2^\prime)\in R_h$, and there exists $q_3\stackrel{a}\Rightarrow_3 q_3^\prime$ such that $(q_2^\prime,q_3^\prime)\in R_h^\prime$ .Hence, $(q_1^\prime,q_3^\prime)\in R_{h+h^\prime}$. Therefore, $R_{h+h^\prime}$ is an $(h+h^\prime)$-approximate weak bisimulation relation of $T_1$ and $T_3$. \hfill $\Box$
\\
\textbf{Proposition 3}: Let $T_1,T_2$ be two nondeterministic LTSs. Then \\
(1). For all $h\in R_0^{+},\epsilon,\delta\in[0,1)$, T$_1\cong_{h,\epsilon,\delta}$T$_1$.\\
\textbf{Proof}: This property is obvious. \hfill  $\Box$
\\
\textbf{Proposition 4}: For any PTS, $\epsilon\in [0,1), \delta\in [0,1), h\in R_0^{+}$.We have \\
(1).$\cup_{i\in I}R_i$ is an approximate probabilistic bisimulation if $R_i$ is an approximate probabilistic bisimulation for any $i\in I$. \\
(2). $\cong_{h,\epsilon,\delta}$ is the largest $(h,\epsilon,\delta)$-approximate probabilistic bisimulation relation. \\
\textbf{Proof}: \\
(2) is implied by (1) and definition 11, so it suffices to prove (1). \\
Let $I$ be an indexing set and $R_i$ an $(h,\epsilon,\delta)$-approximate probabilistic bisimulation for each $i\in I$. It is enough to show that $\cup_{i\in I}R_i$ is an $(h,\epsilon,\delta)$-approximate probabilistic bisimulation relation. \\
Let $(<q_1,\beta,p_{01}>,<q_2,\beta,p_{02}>)\in \cup_{i\in I}R_i$ \\
$\therefore (<q_1,\beta, p_{01}>,<q_2,\beta,p_{02}>)\in R_k$ for some $k\in I$. \\
We proceed by distinguishing between the different kinds of transition relations. \\
Case 1: action step relation:\\
Suppose $<q_1,\beta, p_{01}>\stackrel{\alpha[p_1]}\Longrightarrow_{1} <q_1^\prime,\beta,p_{01}^\prime>,\alpha\in \Sigma\cup\{\tau\}$.\\
$\because(<q_1,\beta,p_{01}>,<q_2,\beta, p_{02}>)\in R_k$, $R_k$ is an $(h,\epsilon,\delta)$-approximate probabilistic bisimulation relation\\
$\therefore \exists q_2^\prime \in Q_2$ such that $<q_2,\beta, p_{02}> \stackrel{\alpha[p_2]}\Longrightarrow_{2}<q_2^\prime,\beta,p_{02}^\prime>$, Pr$(|p_1-p_2|\leqslant \epsilon)\geqslant 1-\delta$, \\
and $(<q_1^\prime,\beta^\prime,p_{01}^\prime>,<q_2^\prime,\beta^\prime,p_{02}^\prime>)\in R_k$ \\
$\therefore (<q_1^\prime,\beta^\prime,p_{01}^\prime>,<q_2^\prime,\beta^\prime,p_{02}^\prime>)\in R_k\subseteq \cup_{i\in I}R_i $  
\\
Case 2: time step relation:\\
Suppose $<q_1,\beta, p_{01}>\stackrel{r,\rho,[1]}\twoheadrightarrow{1} <q_1^\prime,\beta,p_{01}^\prime>$ \\
$\because (<q_1,\beta,p_{01}>,<q_2,\beta,p_{02}>)\in R_k$, $R_k$ is an $(h,\epsilon,\delta)$-approximate probabilistic bisimulation \\
$\therefore \exists q_2^\prime \in Q_2$ such that $<q_2,\beta,p_{02}>\stackrel{r,\rho,[1]}\twoheadrightarrow_{2}<q_2^\prime,\beta^\prime,p_{02}^\prime>$ and $(<q_1^\prime,\beta^\prime,p_{01}^\prime>,\\<q_2^\prime,\beta^\prime,p_{02}^\prime>)\in R_k$ \\ 
$\therefore (<q_1^\prime,\beta^\prime,p_{01}^\prime>,<q_2^\prime,\beta^\prime,p_{02}^\prime>)\in R_k \subseteq \cup_{i\in I}R_i$ 
\\ 
Case 3: discontinuous relation: this case trivally holds \\
In conclusion, $\cup_{i\in I}R_i$ is an $(h,\epsilon,\delta)$-approximate probabilistic bisimulation. \hfill $\Box$
\\
\textbf{Proposition 5}: For any PTS, $\epsilon,\epsilon^\prime\in[0,1),\delta,\delta^\prime\in[0,1),h,h^\prime\in R_0^{+}$.We have \\
(1). If $\epsilon\leqslant \epsilon^\prime$, then $\cong_{h,\epsilon,\delta}\subseteq\cong_{h,\epsilon^\prime,\delta}$. \\
(2). If $\delta\leqslant \delta^\prime$, then $\cong_{h,\epsilon,\delta}\subseteq\cong_{h,\epsilon,\delta^\prime}$. \\
(3). If $h\leqslant h^\prime$, then $\cong_{h,\epsilon,\delta}\subseteq\cong_{h^\prime,\epsilon,\delta}$. \\
\textbf{Proof}: (1) By definition 11, if $\epsilon\leqslant \epsilon^\prime$ then we can immediately conclude that every $(h,\epsilon,\delta)$-approximate probabilistic bisimulation relation is an $(h,\epsilon^\prime,\delta)$-approximate probabilistic bisimulation relation and, hence, (1) holds. Similarly, (2) and (3) holds too. \hfill   $\Box$
\\
\textbf{Lemma 1} (Hoeffding’s inequality \cite{gatsis2019statistical})	\\
Consider a sequence $\{\gamma_k,k=0,1,…,N-1\}$ of independent identically distributed (i.i.d.) random variables taking values in [0,1] with mean $q$. Let $\hat{q}_N=\frac{1}{N}\sum_{k=0}^{N-1}\gamma_k$ be the sample average. Then for any $\epsilon>0$, we have that 
\begin{equation}
P(\hat{q}_N\geqslant q+\epsilon)\leqslant exp\{-2N\epsilon^2\}
\end{equation}
\begin{equation}
P(\hat{q}_N\leqslant q-\epsilon)\leqslant exp\{-2N\epsilon^2\}   
\end{equation}
where the probability is with respect to the random sequence $\{\gamma_k, k=0,1,…,N-1\}$. \\
The result essentially states that there is a low probability that the sample average deviates much from the true probability and further provides an explicit bound on this probability. Note that, inequalities (1) and (2) can be rewrite to inequality (3), which likes the condition of our approximate probabilistic bisimulation.
\begin{equation}
	P(\vert \hat{q}_N-q \vert\leqslant \epsilon)\geqslant 1 - exp\{-2N\epsilon^2\}
\end{equation}
Assume that $q$ and $\hat{q}_N$  are the probabilities of two transitions, in order to ensure inequality (3) satisfy the confidence interval, we only need to ensure inequality (4) holds. 
\begin{equation}
	1-exp\{-2N\epsilon^2\}\geqslant 1-\delta
\end{equation}
In other words, we only need the sample sequence $N$ satisfy inequality (5). That is to say, by comparing the difference of transition probability in $N$ trails, we can decide whether the two probabilistic transitions are approximate probabilistic bisimilar or not. 
\begin{equation}
	N\geqslant \frac{-ln\delta}{2\epsilon^2} 
\end{equation}
As in our definition of approximate probabilistic bisimulation, we can’t decide directly whether two probabilistic transitions are approximate probabilistic bisimilar in an experiment, we need to adopt statistical methods to decide it. 

\section{ Language }
While probabilistic model checking logics takes a big step towards combining performance analysis and model checking techniques, and can express probability related properties such as \textquotedblleft with probability at most 0.05, the system will reach a deadlock state within 10 minutes\textquotedblright, but they are limited to producing only true or false responses, as they are still logics. But in real life, we need to evaluate the system with much more properties besides above, such as what is the minimum(maximum) probability that the system reach a fault state within 10 minutes. 
\\
\indent In this section, we introduce a real valued formal language, towards the unification of model checking and performance evaluation. This language can express not only the properties of states but also of paths, called Continuous Time Real valued Measurement Language (CTRML). 
\\
\textbf{Basic definitions} \\
\textbf{Definition 12 (state formula)}: A CTRML state formula $\phi$ is defined as a function that maps a given state space \textit{S} to real values:
\begin{equation}
	\phi \colon \textit{S}\rightarrow R
\end{equation}
\textbf{Definition 13 (restricted state formula)}: A CTRML state formula $\varphi$ is defined as a function that maps a given state space \textit{S} to the interval [0, 1]:
\begin{equation}
	\varphi \colon \textit{S}\rightarrow  [0, 1]
\end{equation}
\textbf{Definition 14 (path formula)}: A CTRML path formula $\psi$ is defined as a function 
\begin{equation}
	\psi: S^\omega\rightarrow R
\end{equation}
\textbf{Definition 15 (restricted path formula)}: A CTRML restricted path formula $\rho$ is defined as a function that maps from the set of paths to the interval [0, 1]: 
\begin{equation}
	\rho:S^\omega\rightarrow [0, 1]
\end{equation}
\textbf{Syntax} \\
Let $r\in$ AR(atomic restricted state formula), the syntax of CTRML can be defined recursively as follows: 
\\
$\phi\ \colon\colon \ =\ \varphi \ |\ \phi \odot \phi \ |\ $ MIN[$\psi]\ |\ $ MAX[$\psi]$ \\
$\varphi\ \colon\colon \ =\ r\ |\ 0\ |\ 1\ |\ \phi \bowtie c\ |\ \phi \bowtie \phi \ |\ 1-\varphi \ |\ $ MIN[$\rho] \ | \ $ MAX[$\rho]\ |\ $P$_{\sim p}[\rho]$  \\ 
$\psi\ \colon\colon\ =\ $ X$\phi\ |\ \phi \textbf{U}_\odot^{\leqslant t} \phi$ \\ 
$\rho\ \ \colon\colon\ =\ $ X$\varphi\ |\ \varphi \textbf{U}_\times^{\leqslant t} \varphi $ \\
where $\odot \in \{\times,+\}, \bowtie \in \{\geqslant,>,=,<,\leqslant \}, \sim\in \{\geqslant,\leqslant\}, c\in R, p\in [0,1]$. 
\\ 
\textbf{Semantics} \\
Let $f$, $g$ be state formulas, then the semantics of CTRML can be recursively defined as follows: \\
$\cdot$ If $h = f\bowtie c$, then $h(s) = 1$ if $f(s) \bowtie c$ holds, otherwise $h(s)=0$. \\
$\cdot$ If $h = f\bowtie g$, then $h(s) = 1$ if $f(s) \bowtie g(s)$ holds, otherwise $h(s)=0$. \\
$\cdot$ If $h = f + g$, then $h(s) = f(s) + g(s)$.\\
$\cdot$ If $h = f\cdotp g$, then $h(s) = f(s)\cdotp g(s)$. \\
$\cdot$ If $h = 1-f$, then $h(s) = 1- f(s)$. \\
$\cdot$ If $h = $MIN$[\psi]$, then $h(s) = min\ \mu_\psi(S_s^\omega)$. \\
$\cdot$ If $h = $MAX$[\psi]$, then $h(s) = max\ \mu_\psi(S_s^\omega)$. \\
$\cdot$ If $\psi = $X$f$, then $\psi(s_0, s_1, s_2, \ldots )= f(s_1)$. \\
$\cdot$ If $\psi = f\textbf{U}_\odot^{\leqslant t}g$ with $\odot \in \{\times,+\}$ and $t\in N\cup\{\infty\}$, then $\psi(s_0,s_1,s_2,\ldots)$= \\
$(\odot_{i=0}^{j-1}f(s_i))\odot g(s_j)$,if $\exists j:0\leqslant j\leqslant t,g(s_j)\geqslant 0$, and $\forall 0\leqslant i<j,g(s_j)=0$,\\ 
otherwise 0. \\
$\cdot$ If $h = $MIN$[\psi],\psi=$X$\phi$, then $h(s_0) = \min\limits_{\forall \sigma\in Path(s_0,s_1)} \phi(s_1)\cdotp P(s_0,s_1)$ \\
$\cdot$ If $h = $MAX$[\psi],\psi=$X$\phi$, then $h(s_0) = \max\limits_{\forall \sigma\in Path(s_0,s_1)} \phi(s_1)\cdotp P(s_0,s_1)$ \\
$\cdot$ If $h = $MIN$[f\textbf{U}_\odot^{\leqslant t}g]$ with $\odot \in \{\times,+\}$ and $t\in N\cup\{\infty\}$, then $\psi(s_0,s_1,s_2,\ldots)$=\\
$min(\odot_{i=0}^{j-1}f(s_i))\cdotp g(s_j)$, if $\exists j:0\leqslant j\leqslant t,g(s_j)\geqslant 0$, and $\forall 0\leqslant i<j,g(s_j)=0$, \\
otherwise 0. \\
$\cdot$ If $h = $MAX$[f\textbf{U}_\odot^{\leqslant t}g]$ with $\odot \in \{\times,+\}$ and $t\in N\cup\{\infty\}$, then $\psi(s_0,s_1,s_2,\ldots)$=\\
$max(\odot_{i=0}^{j-1}f(s_i))\cdotp g(s_j)$, if $\exists j:0\leqslant j\leqslant t,g(s_j)\geqslant 0$,  and $\forall 0\leqslant i<j,g(s_j)=0$, \\
otherwise 0. \\
$\cdot$ P$_{\sim p}[\rho]$ = $\frac{\mathrm{MAX}[\rho]-p}{\mathrm{MAX}[\rho]-\mathrm{MIN}[\rho]}$ if $\sim=\geqslant$, otherwise $\frac{p-\mathrm{MIN}[\rho]}{\mathrm{MAX}[\rho]-\mathrm{MIN}[\rho]}$.  
\vspace{1ex} \\
Let us recall the definition of Probabilistic bisimulation in \cite{jonsson2001probabilistic}, which is defined by using a  combined transition. Traditionally, the probability property described by PCTL, e.g., P$_{\geqslant0.9}$[True$ \cup^{\leqslant 10}$Fault] can be only true or false, precisely if the probability of the property \textquotedblleft True$\cup^{\leqslant 10}$Fault\textquotedblright \ is lower than 0.9, then the result is false. But, by using the definition of Probabilistic bisimulation which use a combined transition, we can get this probability when the maximum probability is greater than 0.9 and the minimum probability is lower than 0.9. In my new language, we use the same symbolic representation, but use a different semantic interpretation. It describes the degree to which the probability property holds. 
\\
\textbf{Definition 16 (Probabilistic bisimulation) \cite{jonsson2001probabilistic}}: In a PTS, an equivalence relation $R$ over $S$ is a probabilistic bisimulation if $sRt$ implies that whenever $s\stackrel{a}\rightarrow \pi$ for some action $a$ and distribution $\pi$, then there is a distribution $\rho$ such that $t\stackrel{a}\rightarrow \rho$ is a combined transition with $\pi R\rho$.
\\ 
\textbf{Algorithm} 
\\
Below, algorithms for \textbf{U},P,MIN and MAX main operators are given. 
\begin{algorithm}[ht]
	\caption{Compute MIN[$\phi_1 \textbf{U}_\times^{\leqslant t} \phi_2$]}
	\label{alg:Framwork}
	\begin{algorithmic}[1]
		\Require
		$\phi_1 \textbf{U}_\times^{\leqslant t} \phi_2$, start state $s_0$, $c=\infty$, pathSets.
		\State \textbf{for} $path_i$,(i=1,\ldots,|pathSets|)
		\label{code:fram:extract}
		\State \qquad $N=|path_i|$,$val=1,j=0,t_j=0$ 
		\State \qquad \textbf{while} ($j<N$ and $t_j\leqslant t$) \textbf{do}
		\State \qquad \qquad \textbf{if} $s_j\in Sat(\phi_2)$ and $t_j\leqslant t$ \textbf{then}	
		\State \qquad \qquad \qquad $val=1$
		\State \qquad \qquad \textbf{else if} $s_j\in Sat(\phi_1)$ and $s_{j+1}\in Sat(\phi_1)$ and $t_j\leqslant t$ \textbf{then}
		\State  \qquad \qquad \qquad $val=val\cdotp$ P$(s_j,s_{j+1})$
		\State \qquad \qquad \textbf{else if} $s_j\in Sat(\phi_1)$ and $s_{j+1}\in Sat(\phi_2)$ and $t_j\leqslant t$ \textbf{then}	
		\State  \qquad \qquad \qquad $val=val\cdotp$ P$(s_j,s_{j+1})$			
		\State  \qquad \qquad \textbf{else} $val=\infty$
		\State  \qquad \qquad \textbf{end if}
		\State  \qquad \qquad j=j+1	
		\State  \qquad \textbf{end while}		
		\State  \qquad \textbf{if} $val<c$
		\State  \qquad \textbf{then} $c=val$	
		\State  \qquad \textbf{end if} 					
		\State  \textbf{end	for}	
		\State  return $c$						
	\end{algorithmic}
\end{algorithm}
\\
Algorithm for computing MAX[$\phi_1 \textbf{U}_\times^{\leqslant t} \phi_2$] is similar to MIN[$\phi_1 \textbf{U}_\times^{\leqslant t} \phi_2$], we only need to change \textquotedblleft$c=\infty$\textquotedblright \ to  \textquotedblleft$c=-\infty$\textquotedblright \ and line 14 to \textquotedblleft if $val>c$ \textquotedblright \ instead.
\begin{algorithm}[h]
	\caption{Compute P$_{\sim p}[\rho]$}
	\label{alg:Framwork}
	\begin{algorithmic}[1]
		\Require
		$\rho$, pathSets.
		\State \textbf{if} $\sim=\geqslant$ \textbf{then}
		\label{code:fram:extract}
		\State \qquad 
		return $\frac{\mathrm{MAX}[\rho]-p}{\mathrm{MAX}[\rho]-\mathrm{MIN}[\rho]}$
		\label{code:fram:trainbase}
		\State \textbf{else}
		\label{code:fram:add}
		\State \qquad 
		return $\frac{p-\mathrm{MIN}[\rho]}{\mathrm{MAX}[\rho]-\mathrm{MIN}[\rho]}$
		\State  \textbf{end if}
	\end{algorithmic}
\end{algorithm}
\begin{algorithm}[H]
	\caption{Compute MAX[$\phi_1 \textbf{U}_+^{\leqslant t} \phi_2$]}
	\label{alg:Framwork}
	\begin{algorithmic}[1]
		\Require
		$\phi_1 \cup_+^{\leqslant t} \phi_2$, start state $s_0$, $c=-\infty$, pathSets.
		\State \textbf{for} $path_i$,(i=1,\ldots,|pathSets|)
		\label{code:fram:extract}
		\State \qquad $N=|path_i|$,$val=0,j=0,t_j=0$ 
		\State \qquad \textbf{while} ($j<N$ and $t_j\leqslant t$) \textbf{do}
		\State \qquad \qquad \textbf{if} $s_j\in Sat(\phi_2)$ and $t_j\leqslant t$ \textbf{then}	
		\State \qquad \qquad \qquad $val=val+\phi_2(s_j)$
		\State \qquad \qquad \textbf{else if} $s_j\in Sat(\phi_1)$ and $s_{j+1}\in Sat(\phi_1)$ and $t_j\leqslant t$ \textbf{then}
		\State  \qquad \qquad \qquad $val=val+\phi_1(s_j)$
		\State \qquad \qquad \textbf{else if} $s_j\in Sat(\phi_1)$ and $s_{j+1}\in Sat(\phi_2)$ and $t_j\leqslant t$ \textbf{then}	
		\State  \qquad \qquad \qquad $val=val+\phi_2(s_{j+1})$		
		\State  \qquad \qquad \textbf{else} $val=-\infty$
		\State  \qquad \qquad \textbf{end if}
		\State  \qquad \qquad j=j+1	
		\State  \qquad \textbf{end while}		
		\State  \qquad \textbf{if} $val>c$
		\State  \qquad \textbf{then} $c=val$	
		\State  \qquad \textbf{end if} 					
		\State  \textbf{end	for}	
		\State  return $c$						
	\end{algorithmic}
\end{algorithm}

Algorithm for computing MIN[$\phi_1 \textbf{U}_+^{\leqslant t} \phi_2$] is similar to MAX[$\phi_1 \textbf{U}_+^{\leqslant t} \phi_2$], we only need to change “$c=-\infty$” to “$c=\infty$” and line 14 to “if $val<c$” instead.
\begin{algorithm}[H]
	\caption{Compute MIN[X$\varphi$]}
	\label{alg:Framwork}
	\begin{algorithmic}[1]
		\Require
		X$\varphi$, start state $s_0$, $c=\infty$, pathSets.
		\State \textbf{for} $path_i$,(i=1,\ldots,|pathSets|)
		\label{code:fram:extract}
		\State \qquad \textbf{if} $s_1\in Sat(\varphi)$ \textbf{then}
		\State \qquad \qquad $val=\varphi(s_1)\cdot$ P($s_0,s_1)$
		\State \qquad \qquad \textbf{if} $val<c$ \textbf{then}	
		\State  \qquad \qquad \qquad $c=val$
		\State  \qquad \qquad  \textbf{end if}
		\State  \qquad \textbf{else}
		\State  \qquad \qquad  $val=\infty$
		\State  \qquad \textbf{end if}
		\State  \textbf{end	for}	
		\State  return $c$						
	\end{algorithmic}
\end{algorithm}
Algorithm 4 is used for computing the minimal value along paths that satisfy the formula X$\varphi$. MAX[X$\phi$] can be computed the same as MIN[X$\varphi$].

\section{ Case study(Modeling) }
In this section, we provide a case study to illustrate the application of   in the real world. Beforehand, we model a nuclear reactor depicted in Fig. 2, which is concerned with the temperature control. This example is adapted from \cite{bergstra2005process,alur1996automatic} with a slight modification. There are five modes: No\_rod, rod1, rod2, Malf, Deactive. The latter two modes model the case of communication messages lost and the temperature increases without interruption. When the temperature reaches the threshold, the core will be shut down. We take the following informal description of the behavior of the reactor as the starting point of our formal description. 
\\
\indent Initially, the temperature of the reactor core is 510 degree centigrade and both control rods are outside the reactor core. With the control rods outside the reactor core, the temperature $T$ of the reactor core increases according to the differential equation $\dot T= 0.1T - 50$. The reactor must be shut down if the temperature increases beyond 550 degree centigrade. To prevent a shutdown, the reactor will nondeterministic choose to communicate with one of the control rods to put the rod into the reactor core once the temperature becomes 550 degree centigrade, and the nondeterministic choice is determined by a scheduler. The communication process may via wire channel or wireless channel, through which the communication message may be lost with a probability due to external environment or other factors influence. Such systems operate under unpredictable channel conditions following unknown distributions, which are more often observable via a finite amount of collected channel sample measurements \cite{rappaport2015wideband,halperin2010predictable}. When control rod 1 is added into the core, the temperature $T$ decreases slowly according to the differential equation $\dot T= 0.1T - 56$ but with high probability that the communication message will reach control rod 1, assume that the probability is $\frac{9}{10}$. For control rod 2, the temperature $T$ decreases quickly according to the differential equation $\dot T= 0.1T - 60$ but with low probability that the communication messages will reach control rod 2, assume that the probability is $\frac{4}{5}$. The probability is 1 for other transitions without labeling probability. When the communication message is lost, the temperature of the reactor core will increases continuously according to the original differential equation. When the temperature $T$ reaches 580 degree centigrade, the reactor core will shut down in order to keep the reactor core safe.  
\\
\indent An additional requirement asserts that, whenever one of the control rods is removed from the reactor core, it can’t be put back into the core for $c$ seconds, for a time parameter $c$. This requirement is enforced by the clock $y_i (i=1, 2)$, which measures the elapsed time since control rod $i$ has been removed from the core. The control rods synchronize with the core through shared edge labels such as add1. The control rods are modeled on top of Fig. 2. 
\\
\indent The given initial condition of the system is (No\_rod, Out1, Out2)$\wedge$ T=510$\wedge y_1=c\wedge y_2=c$.
\\
\begin{figure}
	\centering
	\includegraphics[width=0.7\linewidth]{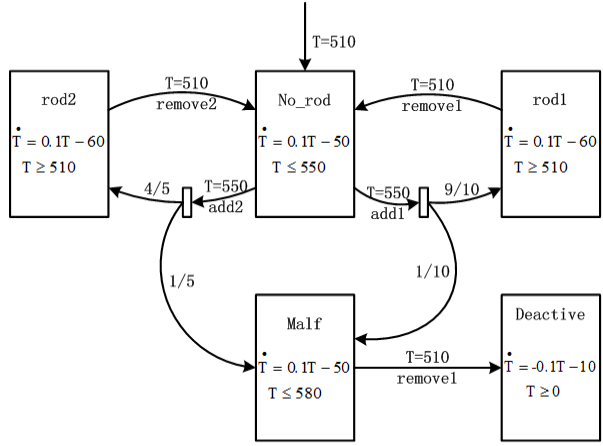}
	\caption{A reactor core with two rods. The rectangular boxes denote random events selecting a transition variant with probabilities denoted along the transition arcs.}
	\label{fig:2}
\end{figure}
\textbf{modeling}
\\
$C=(T=510)\  ^\blacktriangle \hspace{-0.8em} ^\wedge \ C^{out}$\vspace{1ex} 
\\
$C^{out}=(T\leqslant 510\wedge \dot{T}=0.1T-50)\ _\blacktriangledown \hspace{-0.9em} ^\bigcap \hspace{0.05em}$  \vspace{1ex} 
\\
\indent $\left( { \sigma_{rel}^*\left((T=550):\rightarrow \left((T^\cdot=^\cdot \hspace{-0.3em} T)\ _\blacktriangledown \hspace{-1.2em} \sqcap \widetilde{\widetilde{s_1(add)}}\cdot (0.9C^{in1}+0.1D)\right)\right) \vspace{1ex} }\right. \\
\left. { + \sigma_{rel}^*\left((T=550):\rightarrow \left((T^\cdot= ^\cdot \hspace{-0.2em} T)\ _\blacktriangledown \hspace{-1.2em} \sqcap \widetilde{\widetilde{s_2(add)}}\cdot (0.8C^{in2}+0.2D)\right) \right) }\right)$ \vspace{1ex} 
\\
$C^{in1}=\left(T\geqslant 510\wedge \dot{T}=0.1T-56\right)\ _\blacktriangledown \hspace{-0.9em} ^\bigcap \hspace{0.05em}\  \sigma_{rel}^*\left((T=510):\rightarrow \left((T^\cdot= ^\cdot \hspace{-0.2em} T)\ _\blacktriangledown \hspace{-1.2em} \sqcap \widetilde{\widetilde{s_1(rmv)}}\cdot C^{out}\right)\right) $\vspace{1ex} 
\\
$C^{in2}=\left(T\geqslant 510\wedge \dot{T}=0.1T-60\right)\ _\blacktriangledown \hspace{-0.9em} ^\bigcap \hspace{0.05em}\  \sigma_{rel}^*\left((T=510):\rightarrow \left((T^\cdot= ^\cdot \hspace{-0.2em} T)\ _\blacktriangledown \hspace{-1.2em} \sqcap \widetilde{\widetilde{s_2(rmv)}}\cdot C^{out}\right)\right) $\vspace{1ex} 
\\
$D=\left(T\leqslant 580\wedge \dot{T}=0.1T-50\right)\ _\blacktriangledown \hspace{-0.9em} ^\bigcap \hspace{0.05em}\  \sigma_{rel}^*\left((T=510):\rightarrow (T^\cdot= ^\cdot \hspace{-0.2em} T)\ _\blacktriangledown \hspace{-1.2em} \sqcap \widetilde{\widetilde{s_3(shutdown)}}\cdot C^{off}\right) $\vspace{1ex} 
\\
$C^{off}=\left(T\geqslant 0\wedge \dot{T}=-0.15T\right)\ _\blacktriangledown \hspace{-0.9em} ^\bigcap \hspace{0.05em}\  \sigma_{rel}^* \left( (T=0):\rightarrow (T^\cdot= ^\cdot \hspace{-0.2em} T)\ _\blacktriangledown \hspace{-1.2em} \sqcap nil \right) $\vspace{1ex} 
\\
$R_1=\sigma_{rel}^*\left( \widetilde{\widetilde{r_1(add)}}\cdot \sigma_{rel}^* \left( \widetilde{\widetilde{r_1(rmv)}}\cdot \sigma_{rel}^c(R_1) \right) \right) $ \vspace{1ex} 
\\
$R_2=\sigma_{rel}^*\left( \widetilde{\widetilde{r_2(add)}}\cdot \sigma_{rel}^* \left( \widetilde{\widetilde{r_2(rmv)}}\cdot \sigma_{rel}^c(R_2) \right) \right) $ \vspace{1ex} 
\\
$E=\sigma_{rel}^*\left( \widetilde{\widetilde{r_3(shut)}}\cdot nil \right) $ \vspace{1ex} 
\\
\\
$\partial_H(C\parallel R_1 \parallel R_2 \parallel E)$, where \vspace{1ex} 
\\
$H=\{s_i(d)\mid i\in \{1,2,3\},d\in \{add,rmv,shut\}\}
\cup \{r_i(d)\mid i\in \{1,2,3\},d\in \{add,rmv,shut\}\}$.

\section{ Transition System of pACP$^{srt}_{hs}$, discretization, algorithm and performance evaluation }
\subsection{Transferring pACP$^{srt}_{hs}$ to probabilistic transition system}
Given an pACP$^{srt}_{hs}$ process $S$, we can derive a PTS $PT(S) =\ < Q, L, \rightarrow, Q^0, Y_1, Y_2, H_1,\\ 
H_2, H_3>$ from $S$ by the following procedure: \\
- The set of states $Q = (subp(S)\cup\{\surd\})\times V(S)\times D(S)$, where $subp(S)$ is the set of sub-processes of $S$, e.g., $subp(S) = \{S\}\cup subp(P_1)\cup subp(P_2)$ for $S\colon\colon = p_1\tilde{\tilde{\alpha_1}}\ldotp P_1\oplus  p_2\tilde{\tilde{\alpha_2}}\ldotp P_2$,$\surd$ is introduced to represent the terminal process, meaning that the process has terminated, and $V(S) = \{v|v\in Var(S) \rightarrow Val\}$ is the set of evaluations of the variables in $S$, with $Val$ representing the value space of variables. $D(S)= \{ p |p\in (subp(S)\cup {\surd})\rightarrow [0,1]\}$ is the set of probabilities (or distributions) of reaching current states.Without confusion in the context, we often call an evaluation $v$ a (process) state. Given a state $q\in Q$, we will use $fst(q), snd(q)$ and $trd(q)$ to return the first, second and third component of $q$, respectively.
\\
- The label set $L$ corresponds to the actions of pACP$^{srt}_{hs}$, defined as $L=R_0^{+}\cup\{\tau\}\cup \Sigma$, where $d\in R_0^{+}$ stands for the time progress, the silent action $\tau$ represents a discrete internal action of  pACP$^{srt}_{hs}$. Besides, $\Sigma$ stands for discrete actions that $S$ can perform. 
\\
- $Q^0 = \{(S,\nu,\pi_0)| \nu \in V(S),\pi_0 \in D(S)\}$, represents that $S$ has not started to execute,$\nu$ is the initial process state of $S$, and $\pi_0$ is the initial probability (distribution) of $S$. 
\\
- $Y_1 = \overline{Val}$, represents the set of value vectors corresponding to $Var(S)$.
\\
- Given $q\in Q, H_1(q) = vec(snd (q))$, where function $vec$ returns the value vector corresponds to the process state of $q$. 
\\
- $Y_2 = D$, represents the set of probability corresponds to $D(S)$.
\\
- $H_2:Q\times L\times Q \rightarrow [0,1]$, represents the probability of transitions, $H_2(q)(a)(q^\prime)$ returns the probability of a transition labeled by $a$ from state $q$ to state $q^\prime$.
\\
- $H_3(q)=prob(third(q))$, where function $prob$ returns the probability (or distribution) corresponding to the process state of $q$.
\\
- $\rightarrow$ is the transition relation on $S$, which correspond to the semantics of pACP$^{srt}_{hs}$.

\subsection{Discretization}
Since most differential equations do not have explicit solutions, or it may cost a lot of time to solve them. Researchers often use approximate techniques to solve this problem, and discretization of the dynamics is normally given by discrete approximate. There are a range of different discretization methods for ODEs \cite{stoer2013introduction}, in this paper we also use Euler method the same as in \cite{yan2016approximate}. As discretization is not the focuses of this paper, we do not introduce it in detail here, we refer readers to \cite{yan2016approximate,stoer2013introduction} for more detail information.

\subsection{Algorithm}
As in our definition of approximate probabilistic bisimulation, we define the probability of the transition probability discrepancy beyond a tolerance $\varepsilon$ with a confidence at least $1-\delta$, which is similar to Chebyshev's theorem \cite{gatsis2019statistical}. So, in our algorithm, we need to take statistical methods to decide whether the probabilities of two transitions are bisimilar or not.
\\
\indent Algorithm 1 decides whether $P_1$ and $P_2$ are $(h,\epsilon,\delta)$-approximately probabilistic bisimilar. When $P_1\cong_{(h,\epsilon,\delta)} P_2$, it returns true, otherwise, it returns false. Let $d$ be the discretized time step. The algorithm is then taken in two steps. The first step (lines 1–6) constructs the transition systems for $P_1$ and $P_2$ with time step $d$. $T(P_m).Q$ and $T(P_m).T$ represent the reachable set of states and transitions of $P_m (m=1,2)$, respectively. The second step (lines 7–25) decides whether the transition systems for $P_1$ and $P_2$ are approximately probabilistic bisimilar with the given precisions, especially, (lines 11–16) are used to count the number of times that satisfy the probability difference in $N$ trials.
\\
\begin{algorithm}[htb]
	\caption{Deceding approximate probabilistic bisimulation between two pACP$^{srt}_{hs}$.}
	\label{alg:Framwork}
	\begin{algorithmic}[1]
		\Require
		process P$_1$,P$_2$, the initial state $\beta$, the initial distribution $\pi_0$, the time step $d$, and precision $h,\epsilon,\delta$.
		\Ensure
		T(P$_m)\ldotp$ Q$^0$=\{(P$_m,\beta,\pi_0$)\},T(P$_m$).T$^0$=$\varnothing$ for m=1,2; i=0;
		\State \textbf{repeat}
		\label{code:fram:extract}
		\State \qquad T(P$_m$).T$^{i+1}$=T(P$_m$).T$^i\cup\{q\stackrel{l[p]}\twoheadrightarrow q^\prime|\forall q\in $T(P$_m$).Q$^i$,if$(\exists l\in\{\tau\}\cup\Sigma,q\stackrel{l[p]}\twoheadrightarrow q^\prime)\}\cup$  \qquad \qquad $\{q\stackrel{l,\rho,[p]}\twoheadrightarrow q^\prime|\forall q\in $T(P$_m$).Q$^i$, if$(\exists l=d,q\stackrel{l,\rho,[p]}\twoheadrightarrow q^\prime)$ or $(\exists l=d^\prime,l<d\wedge q\stackrel{l,\rho,[p]}\twoheadrightarrow q^\prime) \wedge$ not$\ (q\stackrel{d^{"},\rho,[p]}\twoheadrightarrow q^\prime)$ for any $d\textacutedbl$ in $(d^\prime,d]$ ) and $snd(q^\prime)(t_j^m)<T_j^m \}$ 
		\label{code:fram:trainbase}
		\State \qquad T(P$_m$).Q$^{i+1}$=T(P$_m$).Q$^i\cup postState$(T(P$_m$).T$^{i+1})$;
		\label{code:fram:add}
		\State \qquad $i$\textleftarrow$ i+1$;
		\State \textbf{until} T(P$_m$).T$^i$ = T(P$_m$).T$^{i-1}$;
		\State T(P$_m$).Q = T(P$_m$).Q$^i$;\ T(P$_m$).T = T(P$_m$).T$^i$;
		\State $B_{h,\epsilon,\delta}^0=\{(q_1,q_2)\in $T(P$_1$).Q$\times$ T(P$_2$).Q|$d($H$_1^1(q_1)$,H$_1^2(q_2))\leqslant h$,d(H$_3^1(q_1)$,H$_3^2(q_2))\leqslant\epsilon\};i=0$
		\State \textbf{repeat}
		\State \qquad k=0; m=0;
		\State \qquad $(q_1,q_2)\in $T(P$_1$).Q$\times$ T(P$_2$).Q$\backslash B_{h,\epsilon,\delta}^i$
		\State \qquad \textbf{do}
		\State \qquad \qquad \textbf{if}( ($\forall q_1\stackrel{l,[p_1]}\twoheadrightarrow q_1^\prime \in$ T(P$_1$).T, then $\exists q_2\stackrel{l,[p_2]}\Longrightarrow q_2^\prime \in $T(P$_2$).T s.t. $(q_1^\prime,q_2^\prime)\in B_{h,\epsilon,\delta}^i$ and $|p_1-p_2| \leqslant \epsilon)$ and ($\forall q_2\stackrel{l,[p_1]}\twoheadrightarrow q_2^\prime \in $T(P$_2$).T, then $\exists q_1\stackrel{l,[p_2]}\Longrightarrow q_1^\prime \in $T(P$_1$).T s.t. ($q_1^\prime,q_2^\prime)\in B_{h,\epsilon,\delta}^i$ and $|p_1-p_2|\leqslant \epsilon$) ) or ( ( $\forall q_1\stackrel{l,\rho,[p_1]}\twoheadrightarrow q_1^\prime \in$ T(P$_1$).T, then $\exists q_2\stackrel{l,\rho,[p_1]}\Longrightarrow q_2^\prime \in $T(P$_2$).T s.t. $(q_1^\prime,q_2^\prime)\in B_{h,\epsilon,\delta}^i)$ and ($\forall q_2\stackrel{l,\rho,[p_1]}\twoheadrightarrow q_2^\prime \in$ T(P$_2$).T, then $\exists q_1\stackrel{l,\rho,[p_1]}\Longrightarrow q_1^\prime \in $T(P$_1$).T s.t. $(q_1^\prime,q_2^\prime)\in B_{h,\epsilon,\delta}^i$) )
		\State \qquad \qquad \textbf{then} $k=k+1; m=m+1$
		\State \qquad \qquad \textbf{else} $k=k+1$
		\State \qquad \qquad \textbf{end if}
		\State \qquad \textbf{while} $k<N$
		\State \qquad \quad \textbf{if} $\frac{m}{N}\geqslant 1-\delta$ 					
		\State \qquad \quad \textbf{then} $B_{h,\epsilon,\delta}^{i+1}=B_{h,\epsilon,\delta}^i \cup \{(q_1,q_2)\}$
		\State \qquad \quad \textbf{end if}
		\State \textbf{until} $B_{h,\epsilon,\delta}^i=B_{h,\epsilon,\delta}^{i-1}$;
		\State $B_{h,\epsilon,\delta}=B_{h,\epsilon,\delta}^{i}$;
		\State \textbf{if} ((P$_1,\beta,\pi_0$),(P$_2,\beta,\pi_0))\in B_{h,\epsilon,\delta}$ \textbf{then}
		\State \qquad return true;
		\State \textbf{else}	
		\State \qquad return false;					
		\State  \textbf{end if}
	\end{algorithmic}
\end{algorithm}

\subsection{Performance evaluation}
Above, we give the language that used to describe the property of the system. Here, we will give some property for performance evaluation and  the numerical results for each query are listed in Table 5 below.\\
(1). What is the minimal probability that the system will eventually reach Malf state (i.e., the temperature above 550 degree) before 60 time units. The query can then be expressed as
\begin{center}
	MIN[ True $\textbf{U}_\times^{\leqslant 60}(T>550)$ ] 
\end{center}
(2). What is the maximal probability that the system will eventually reach Malf state before 60 time units. The query can then be expressed as
\begin{center}
	MAX[ True $\textbf{U}_\times^{\leqslant 60}(T>550)$ ] 
\end{center}
(3). What is the possibility that the probability of the system will eventually reach Malf state before 60 time units is great than or equal to 0.05. The query can then be expressed as
\begin{center}
	P$_{\geqslant 0.05}$[ True $\textbf{U}_\times^{\leqslant 60}$ Malf ] 
\end{center}
(4). What is the possibility that the probability of the system will eventually reach Malf state before 60 time units is great than or equal to 0.1. The query can then be expressed as
\begin{center}
	P$_{\geqslant 0.1}$[ True $\textbf{U}_\times^{\leqslant 60}$ Malf ]
\end{center}
(5). What is the possibility that the probability of the system will eventually reach Malf state before 60 time units is great than or equal to 0.28. The query can then be expressed as
\begin{center}
	P$_{\geqslant 0.28}$[ True $\textbf{U}_\times^{\leqslant 60}$ Malf ]
\end{center} 
(6). What is the possibility that the probability of the system will eventually reach Malf state before 60 time units is less than or equal to 0.5. The query can then be expressed as
\begin{center}
	P$_{\leqslant 0.5}$[ True $\textbf{U}_\times^{\leqslant 60}$ Malf ]
\end{center}
(7). What is the possibility that the probability of the system will eventually reach Malf state before 60 time units is less than or equal to 0.14. The query can then be expressed as
\begin{center}
	P$_{\leqslant 0.14}$[ True $\textbf{U}_\times^{\leqslant 60}$ Malf ]
\end{center}
(8). What is the possibility that the probability of the system will eventually reach Malf state before 60 time units is less than or equal to 0.05. The query can then be expressed as
\begin{center}
	P$_{\leqslant 0.05}$[ True $\textbf{U}_\times^{\leqslant 60}$ Malf ]
\end{center}
(9). What is the maximal cooling medium consumption that the system will eventually reach Malf state before 60 time units. Assume the cooling medium consumption is 100ml per unit of time for control rod1, and 300ml per unit of time for control rod2, (the cooling medium consumption is 0 when control rod1 (or rod2) is not working). The query can then be expressed as
\begin{center}
	MAX[ True $\textbf{U}_+^{\leqslant 60}(T>550)$ ] 
\end{center}

Table 5  Numberical Results for the CTRML Queries \\
\rule[5pt]{11cm}{0.05em}\\  
\textbf{$\sharp$} \qquad \qquad \textbf{Query} \qquad \qquad \qquad  \ \  \textbf{Initial State} \quad \textbf{Numerical Result} \\
\rule[5pt]{11cm}{0.05em}\\  
1 \qquad MIN[ True $\textbf{U}_\times^{\leqslant 60}$ Malf ] \qquad \ \  No$\_$rod \qquad \qquad \quad 0.08 \\
2 \qquad MAX[ True $\textbf{U}_\times^{\leqslant 60}$ Malf ] \qquad \ No$\_$rod \qquad \qquad \quad 0.2  \\ 
3 \qquad P$_{\geqslant 0.05}$[ True $\textbf{U}_\times^{\leqslant 60}$ Malf ] \quad \ \, No$\_$rod \qquad \qquad \quad 1.25 \\ 
4 \qquad P$_{\geqslant 0.1}$[ True $\textbf{U}_\times^{\leqslant 60}$ Malf ] \qquad \ No$\_$rod \qquad \qquad \quad 0.83 \\ 
5 \qquad P$_{\geqslant 0.28}$[ True $\textbf{U}_\times^{\leqslant 60}$ Malf ] \quad \ \, No$\_$rod \qquad \qquad \quad -0.67 \\ 
6 \qquad P$_{\leqslant 0.5}$[ True $\textbf{U}_\times^{\leqslant 60}$ Malf ] \qquad \ No$\_$rod \qquad \qquad \quad \   3.5 \\ 
7 \qquad P$_{\leqslant 0.14}$[ True $\textbf{U}_\times^{\leqslant 60}$ Malf ] \quad \ \, No$\_$rod \qquad \qquad \quad \ 0.5 \\ 
8 \qquad P$_{\leqslant 0.05}$[ True $\textbf{U}_\times^{\leqslant 60}$ Malf ] \quad \ \, No$\_$rod \qquad \qquad \quad -0.25 \\
9 \qquad MAX[ True $\textbf{U}_+^{\leqslant 60}$ Malf ]  \qquad \  No$\_$rod \qquad \qquad \  3526.71ml
\\
\rule[-1pt]{11cm}{0.05em}\\  
\\

\section{Conclusion}
Approximate bisimulation relation is a useful notion for analyzing complex dynamic systems via simpler abstract systems. Existing approximate probabilistic bisimulation  relation consider the difference of the transition probability of two systems within a tolerance range in each experiment. But in real life, this constrains are too restrictive and not robust. On the one hand, the experiment may be influenced by external factors, such as the stability of a dynamical system, it can only satisfy the constrains in most cases. On the other hand, the values of state variables are based on observations, which may be influenced by sensor noise or other perturbations. For the above reasons, we modify the constraints on transition probability to relaxed constraints, which allows for the probability of the transition probability discrepancy beyond a tolerance $\epsilon$ with a confidence at least $1-\delta$, this modification is more reasonable. In this paper, we extend hybrid process algebra with probability, and define a new approximate probabilistic bisimulation relation of hybrid systems modelled by  pACP$^{srt}_{hs}$, and present an algorithm for deciding whether two pACP$^{srt}_{hs}$  processes are approximately bisimilar. At the end, we illustrate our method by presenting an example of nuclear reactor. 
\\
\indent Regarding future work, we will focus on the implementation, on modelling hybrid systems by using hybrid process algebra. Moreover, in this paper, we only give the semantic rules of the proposed probabilistic hybrid process algebra, we do not give any axioms, so this is another research direction.
\begin{acknowledgements}
This work was supported in part by the Aviation Science
Foundation of China under Grant 20185152035 and
Grant 20150652008, in part by the National Natural
Science Foundation of China under Grant 61572253, in
part by the Fundamental Research Funds for the Central
Universities (NJ2019010, NJ20170007, NJ2020022).
We also thank the anonymous referees for their constructive feedback.
\end{acknowledgements}

\bibliographystyle{unsrt}
\bibliography{ref}                

\end{document}